# Moving mountains and white spots of Ceres

Alex Soumbatov Gur

Among surprising results of NASA's Dawn spacecraft mission at dwarf planet Ceres are enigmatic bright spots inside Occator crater and a lone tall pyramid mountain. Here we explain the appearance of such spots, weird mountains, and concomitant craters on celestial bodies. The phenomenon is as follows. A mountain is ejected rotated out of mechanically stressed crust as a result of explosive conical cleavage all over its buried surface. Then the mountain lands upside down i.e. summit up. It may save its integrity or split into parts to form mountains of smaller sizes and/or block debris. The process is a volcanic caldera formation regime. We infer that it is the result of crack expansion and fracture of planetary crust due to inherent stresses. We scrutinize the processes' phenomenology on examples of Ceres' relief features. Pile-ups of bright substances are also synthesized in the course of the newly described phenomenon of explosive ejective orogenesis. Our analyses allow establish a viewpoint that is consistent with a broad range of observational data.

# Contents

# 1 Introduction

In spring 2015 cosmic probe Dawn approached dwarf planet Ceres, the largest asteroid ca. 945km in diameter. In December 2015 after high altitude flybys the probe moved to the lowest of its orbits 385km above the asteroid. Planned research program has successfully ended in June 2016. Now the probe is conducting extended mission.

The results of all orbital series are outstanding. New findings puzzled astronomy community and brought about heated laymen discussions all over the world.

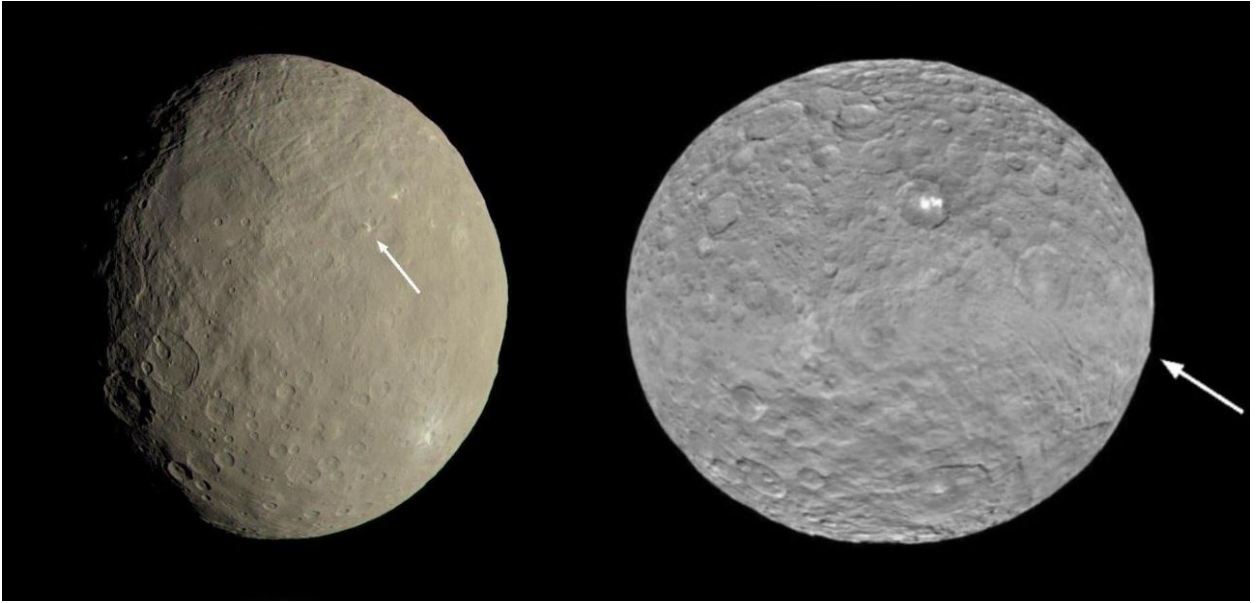

*Fig.1. Two Dawn's images of Ceres. Left: Planet Ceres in colors as human eye would see it. Right: Oriented (north is up) side view of Ceres with Occator crater and its two bright spots. Ahuna mons is pointed by white arrows.*
*(www.photojournal.jpl.nasa.gov./catalog/pia21079)*
*(www.photojournal.jpl.nasa.gov./catalog/pia19619).*

From a distance Ceres looks like a clay sphere with sparse white spots which are randomly spread over its surface (fig.1 (left)). In the most concentrated form of pile-ups brighter substances are accumulated inside 80km wide Occator crater (e.g. fig.1 (right)), that diurnally produces the haze of unknown nature [1, 2]. In Hubble space telescope brighter spots look like diffuse white areas [3]. Data of Hershel telescope imply that Occator is one of the main water vapor exhaust regions of Ceres [4], the other being the region of Dantu crater. The nature of weird spots is thought of by researches as one of the most enigmatic problems of Ceres.

Another mystery is the origin of a lone bright mountain. It is four kilometers high and has 40º slopes. Ahuna Mons is unusually tall for Ceres. The region of this outstanding relief feature with nearby darker craters of comparable sizes is shown in fig.2. One is impossible not to notice the surprising geometrical similarity of the mountain and the nearest large crater. The resemblance stays unexplainable till now.

In this work we are making an attempt to solve above mentioned problems from a heuristic unified point of view.

# 2 Observational misinterpretations

Before the consideration of above mentioned mysteries it is necessary to point out some observational effects. It is not easy to choose which one is crater or mountain just by looking at the image (fig.2). Those beauties are really in the eyes of a beholder. If a viewer rotates the figure he may even switch between crater and mountain impressions. Human vision has no ability to solve the problem without additional information. Shadows are the keys. It is necessary to know the solar position to separate surface ups from downs by the directions of shadows.

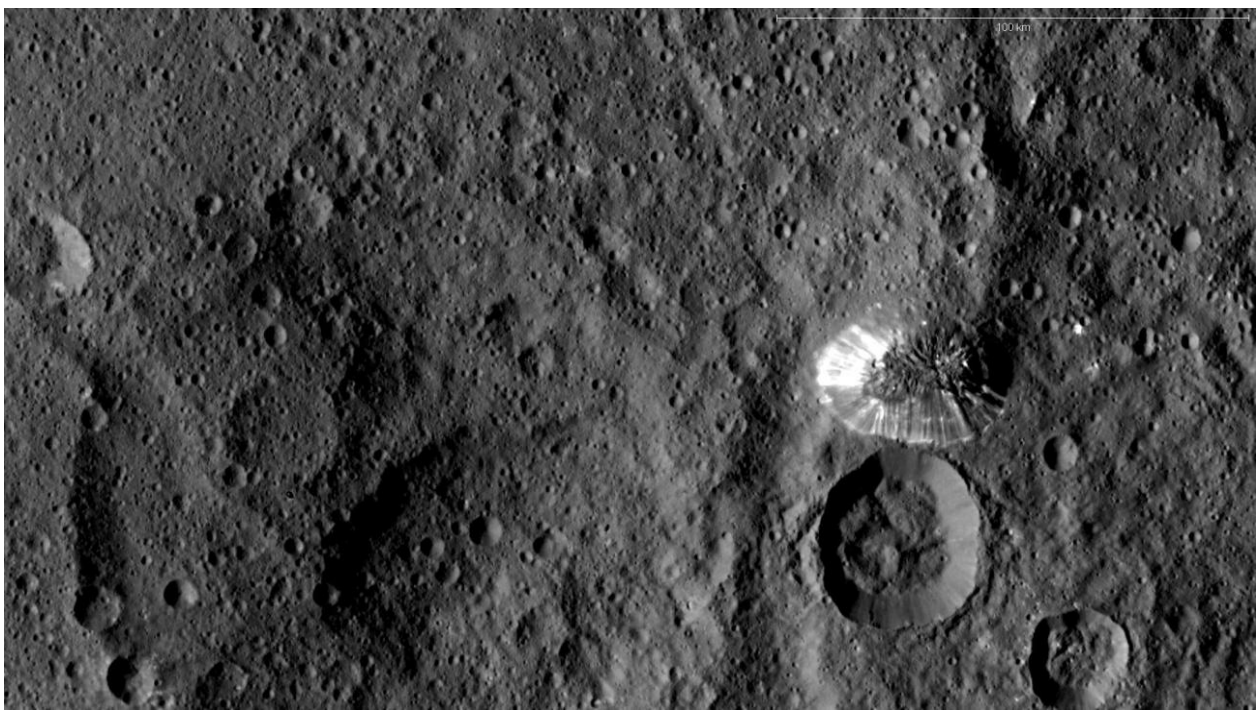

*Fig.2. Image of Ahuna region acquired from the orbit 1470km above the surface of Ceres. Frame area is approx.150 by 270km.*
*(www.photojournal.jpl.nasa.gov./catalog/pia19631)*

By the way, the easy practical rule to find lighting line along the surface (incidence direction or the opposite one) is to analyze black and white relief contrast of small pits and elevations. The computing method is to determine the reciprocal lattice vector for image Fourier expansion on scales of these objects.

According to Dawn's team in the situation depicted in fig.2 sunlight falls from the left to illuminate the bright left slope of Ahuna mons, so the opposite slope of the highest Ceres' mountain is fully shadowed.

All Dawn's data on surface relief depend upon complex photometric three dimensional algorithms and computer programs, elaborated to acquire views and maps of celestial objects. They utilize modeled light reflection parameters and thousands of images, shut from different orbital points in different directions and shadow conditions. Suppose that the photometric complexity leads to a mistake.

It is the mistake of inversion. We found out that the mountain is indeed the crater (and vice versa). In the following the proof of the conclusion is presented and far reaching consequences are drawn out.

# 3 Phenomenology of Ceres' relief formation

## 3.1 Relief of Ahuna region

We outline the positions of objects of our interest in fig.3 with ellipses and circles. For the sake of easy reference we call them crater, mountain, hill, and hillock. For picture preparation we made use of ImageJ, Gwyddion, and IrfanView freeware computer programs. To discern relief details the reader is recommended to sometimes invert the colors of images.

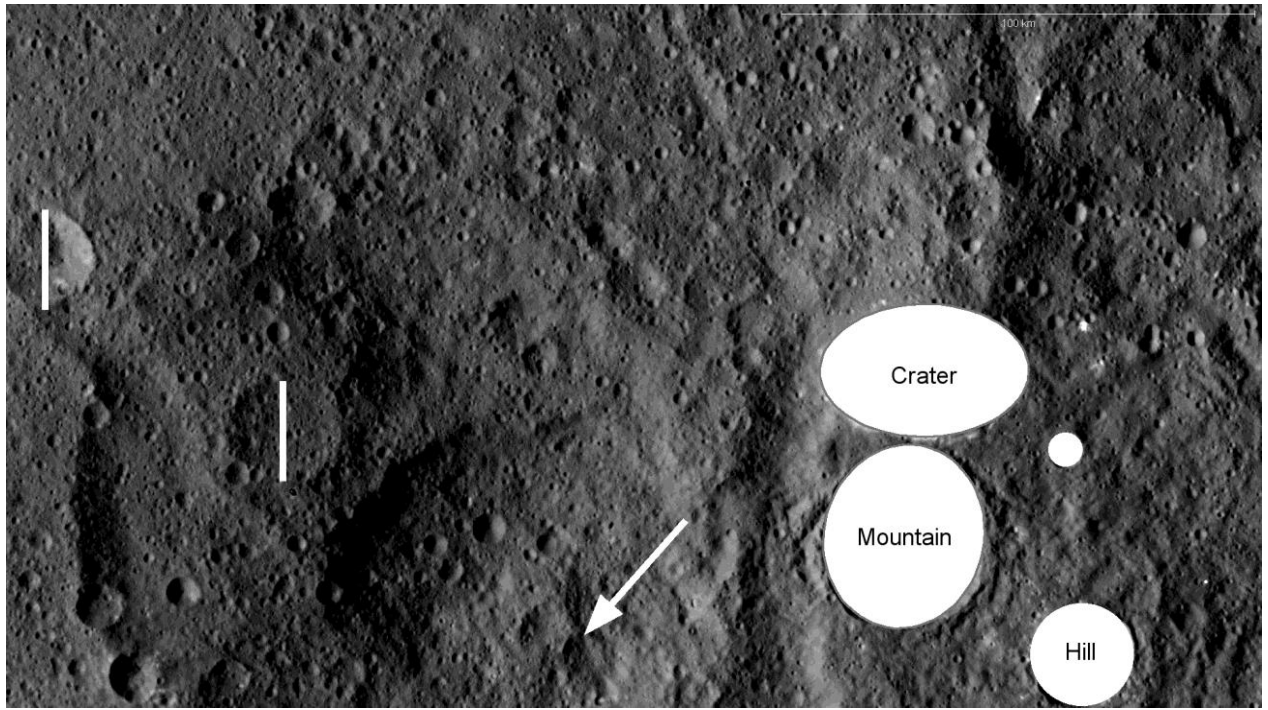

*Fig.3. Figure 2 with the objects of interest masked. The upper ellipse is the crater. The lower one is the table mountain. The big circle borders the position of the smaller table mountain. We refer to it as the hill. The small circle masks coned mountain referred to as the hillock. Two white vertical lines of the same size are shown to compare the dimensions of other round shaped relief objects. White arrow approximately depicts northern direction.*

The objects in question and their close vicinities are shown in fig.4. Opposing the original image description we consider that sunlight falls from the right hand side of figures 2,3,4. Therefore left slopes of the mountain, hill, and hillock look darker. The crater left slope is the brightest of its other slopes. They are also not very dark except for the part of the right slope.

All crater slopes consist of bright streaks directed down approximately perpendicular to the rim. Some of them end with round shaped headings of brighter than average substance. The mountain's and hill's slopes are also lined the same way but look darker and less contrasted. The crater bottom is lined with grooves as well. Small brighter color boulders are dispersed all over the image area, especially to the right and up of the crater.

The elliptical crater is elongated in one of the directions of global fractures (see 3.7), which is shown by white arrow in fig.1 (left). The direction is diagonal to cerean equator.

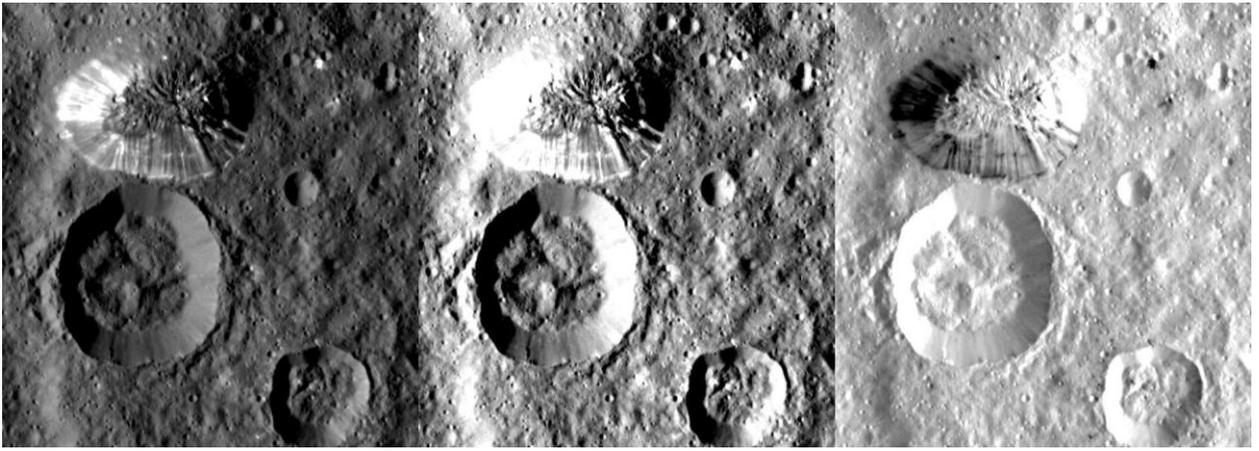

*Fig.4. Left image is the crop from lower right hand part of fig.2. Middle image is the left one with changed contrast. Right image is the negative of the left image.*

The whole appearance of relief around objects of our interest proves the existence of local crustal fractures, running approximately diagonally to all image frames (figs.2,3,4,5) and forming rhomb like net. Its zigzag appearance is especially clear in fig.5 (right). The net is possible to make more discernable e.g. by applying artificial colors to our black and white pictures. Fracture directions are also shown with short white arrows in fig.11, where diagonal fracture net is clearly seen, especially in right hand part of the figure.

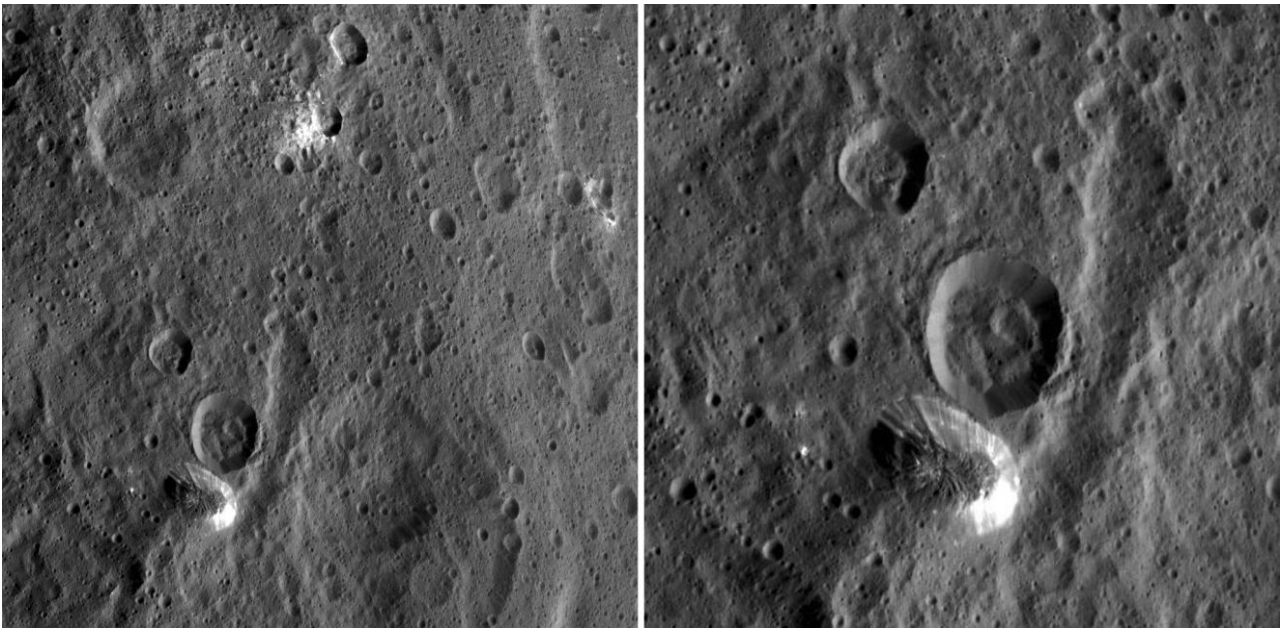

*Fig.5. Left image is the original view. Right image is the contrasted crop from it. (www.photojournal.jpl.nasa.gov./catalog/pia21907)*

Fractures' existence means that local crust was (is) under shear and tensile stresses. The corners of rhombs, or by other words the knot points of crustal lineament net, are stress concentrators. In terms of linear fracture mechanics this means that it is the vicinity of those points, where local stresses strongly rise up in case of external stress application. On Earth all crustal discontinuities are stress concentrators as well as impurities' attractors and channels of diverse fluids.

## 3.2 Mountain-crater ejective birth

First of all let us study relative positions and sizes of our objects. The mountain major axis is almost perpendicular to the one of the crater. Sides of the objects are not destroyed. This means they are very fresh from geologic point of view. It is this fact that allows us to understand their origin.

The overall mountain shape is like that of the crater. Areas restricted by the mountain and the crater are approximately equal. They differ not more than ten percents. The reasons of inequality will be discussed further (see 5.1). So the working hypothesis is that mountain was some way "took out" of the planetary crust, turned over, and "put" nearby. To prove the scenario we are to find coincident features of both objects. Because of their somewhat size differences the features are not to be exactly the same. Their resemblance should be topological.

Let us determine mutual coincidence of crater and mountain sides. The comparison of changing widths of crater and mountain slopes (fig.4) may imply that the upper crater slope coincides to the left mountain one. Now we imagine the initial mountain position inside crater and compare some large relief features of the objects (fig.6).

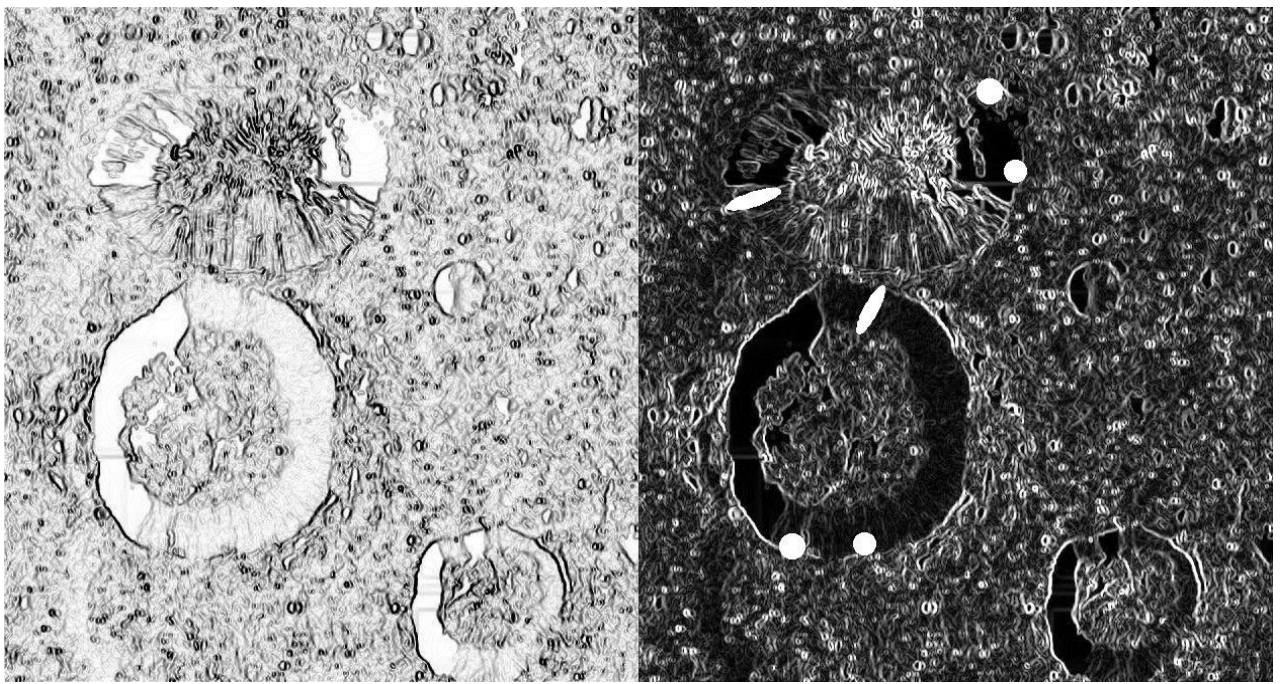

*Fig.6. Processed crop from fig.2. Left: image shows the borders of maximal contrast (edge detection algorithm). Right image is the negative of the left one. To imagine crater-mountain coincidence approx. points of coherence are shown by ellipses and circles (the mountain is to be flipped).*

Notice curved relief lines sectioning the mountain approximately in vertical direction and the crater – in horizontal one (in relation to fig.6 frames). This discontinuity line divides the objects approx. by half. It is the remnant of the crack earlier formed perpendicular to crustal surface. The crack initiates crater formation (see 5.3). The mountain inherits it. At the crater bottom the line is branching (in the region of long white arrow tip in fig.11).

So we conclude that to correctly juxtapose crater and mountain views for direct feature comparison mountain is to be rotated 90º anticlockwise in relation to crater and flipped in vertical direction (see fig.6). Further we make use of NASA's pia20348 image of Ahuna region (fig.14) with the resolution approx. twice better than that of pia19631 (fig.2). The crater and the mountain crops are posed nearby in figs 7,8.

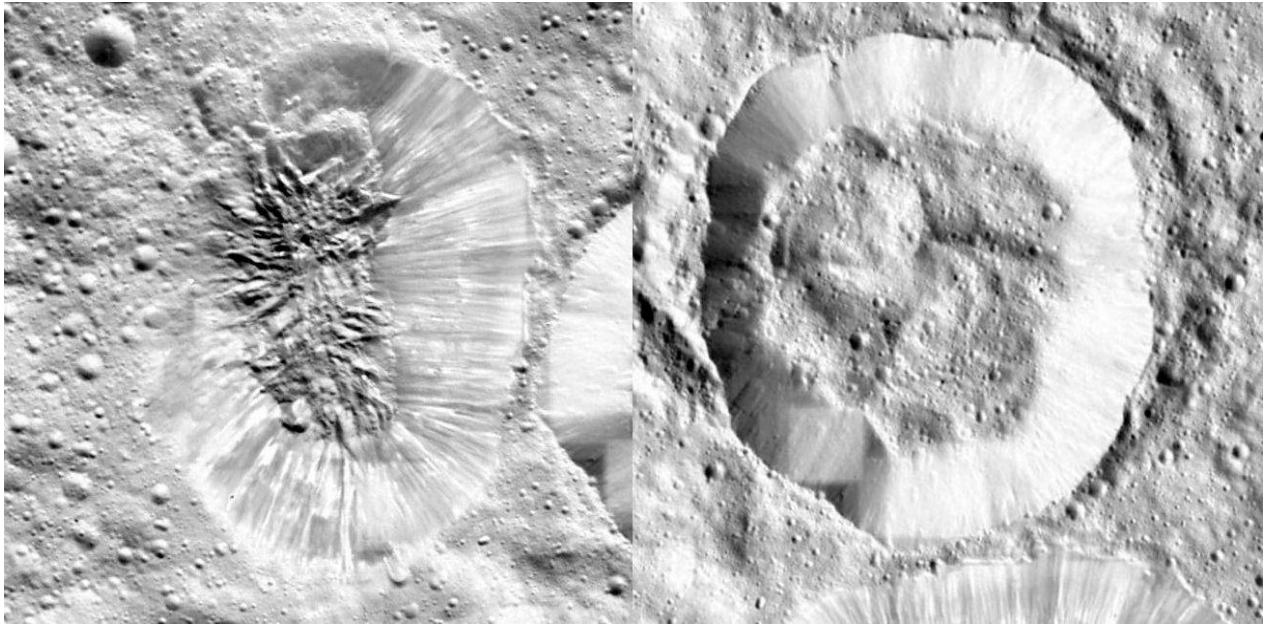

*Fig.7. Images are crops from the NASA's pia20348 image (fig.14). Left one is the crater, right one is the mountain. The backgrounds of both crops were subtracted and their contrasts were enhanced. Bright rectangles on mountain slopes at the 6 to 8 o'clock positions are the results of initial image owners' corrections. (http://www.jpl.nasa.gov/spaceimages/details.php?id=PIA20348)*

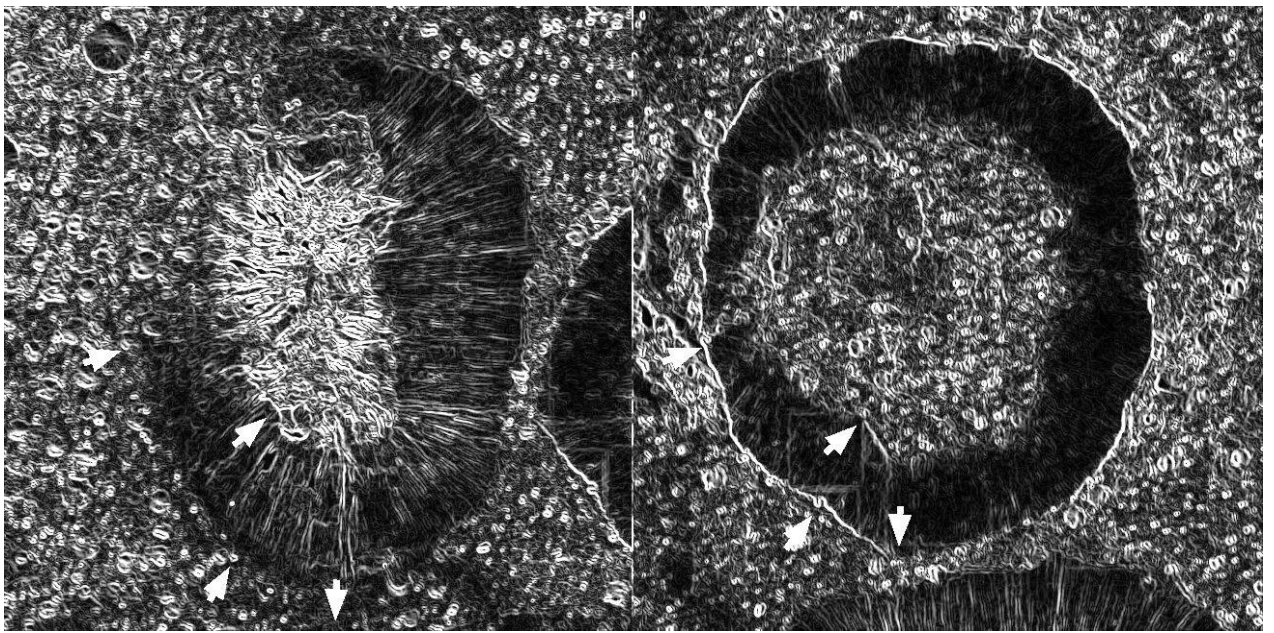

*Fig.8. Processed fig.7. Edge detection algorithm was applied to the image to show the borders of maximal contrast. Arrows nail some coincident perimeter features of the crater and the mountain.*

One can see that borders of the crater and the mountain averagely repeat each other. The same holds for the curvature locations, e.g. marked with white arrows in fig.8. Similarity of the perimeters is obvious in spite of the fact that the crater has a bit longer major axis and shorter minor one.

Now compare the shape of mountain table and the relief of crater bottom (fig.9). Slight contrast line at its slopes (cf. fig.7) allows easy realize the task. The line crosses crater slope streaks approximately perpendicular and sometimes is the place of their continuity breaking. It also determines the positions of scarps. The line appears to mark streak crossings with horizontal crustal layers' borders. The topologic resemblance of the line and the mountain table perimeter is clear. This is one of the crucial proves of mountain ejection.

Several coincident features of crater bottom and mountain table are also pointed in fig.9. An attentive viewer is able to find lots more details of crater-mountain coincidence.

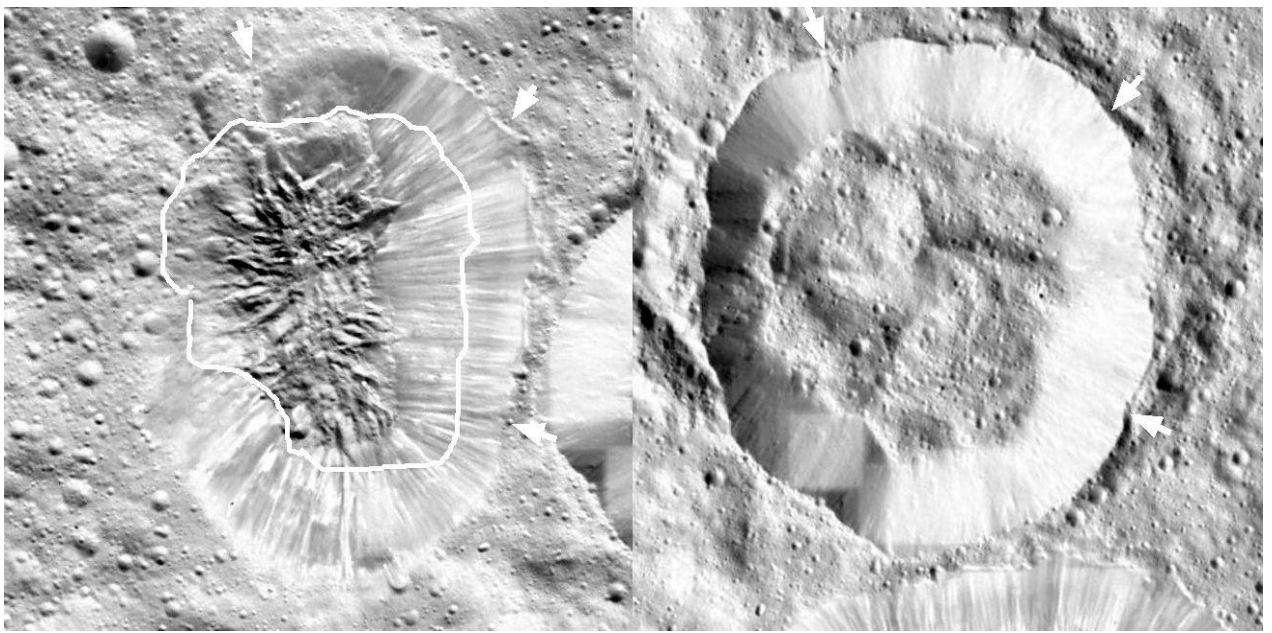

*Fig.9. Marked fig.7. White line depicts the slight relief contrast feature of crater slopes. Arrows point to some coincident features of the objects.*

In fig.10 we made the sizes of both objects approx. equal by rescaling crater image. Its size was multiplied by 0.94 along major crater axis, and by 1.17 - along the minor one. Some features of topological coincidence are shown by white arrows.

Straight white line at the crater slope and curved line at the mountain slope pointed by the lowest arrows are counterparts (fig.10). Comparison of them proves mountain rotation at the start. The outer rim end of the curved line lags behind the inner end. The reason is that the more is the linear velocity of a point of the starting mountain the higher is the plastic friction.

The hillock's cone also has curved streak line of the same nature (e.g. figs.11,14). Features of the kind prove viscous modifications of mountains in the course of their orogenesis. They are genetic marks of rotation gained at moments of origin.

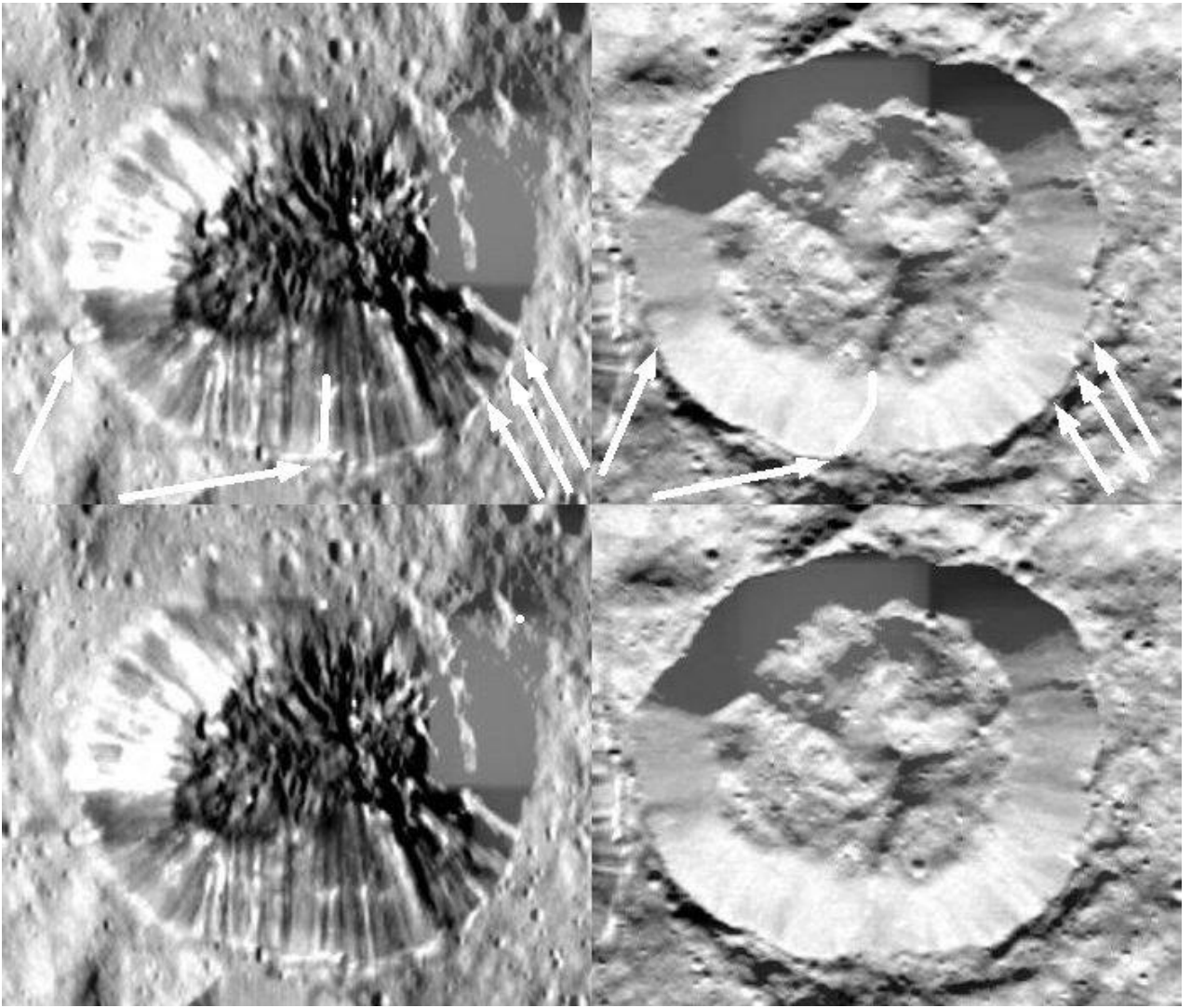

*Fig.10. Crater and mountain crops from fig.2 scaled to approx. same sizes and contrasted. Left photos show the crater, right photos do the mountain 90º counterclockwise rotated and vertically flipped. Some details of coincidence are marked by arrows. Straight and curved white lines pointed by the lowest arrows prove the viscous rotation of the mountain at the start (see text).*

White crater streaks stayed straight while ejection. This means they are stronger than the darker mountain ones. In perpendicular directions concentric scarps of brighter materials are formed. As it was pointed out their curvature marks the section of the conical crater by crustal layer borders i.e. the surfaces of stress concentration and cleavage. That is the reason of specific scarp shape on the right crater slope under the diffuse rectangle (figs.10,11). By the way, the shape of the scarp means impossibility of clockwise mountain movement at the start. Visualized by brighter substances the scarp and the streak kink locations produce a latch against rotation in this direction.

Regional relief around our objects of study is clearly seen in fig.11. Because of landing mountain slowing down its immediate surroundings appear to corrugate. The wavy concentric region in mountain's vicinity is the widest in ca. hill's direction, implying the one is that of losing momentum. The rotational slowing down possibly led to the enhancement of crustal rhombic relief fractures around the mountain, shown by short white arrows in the figure.

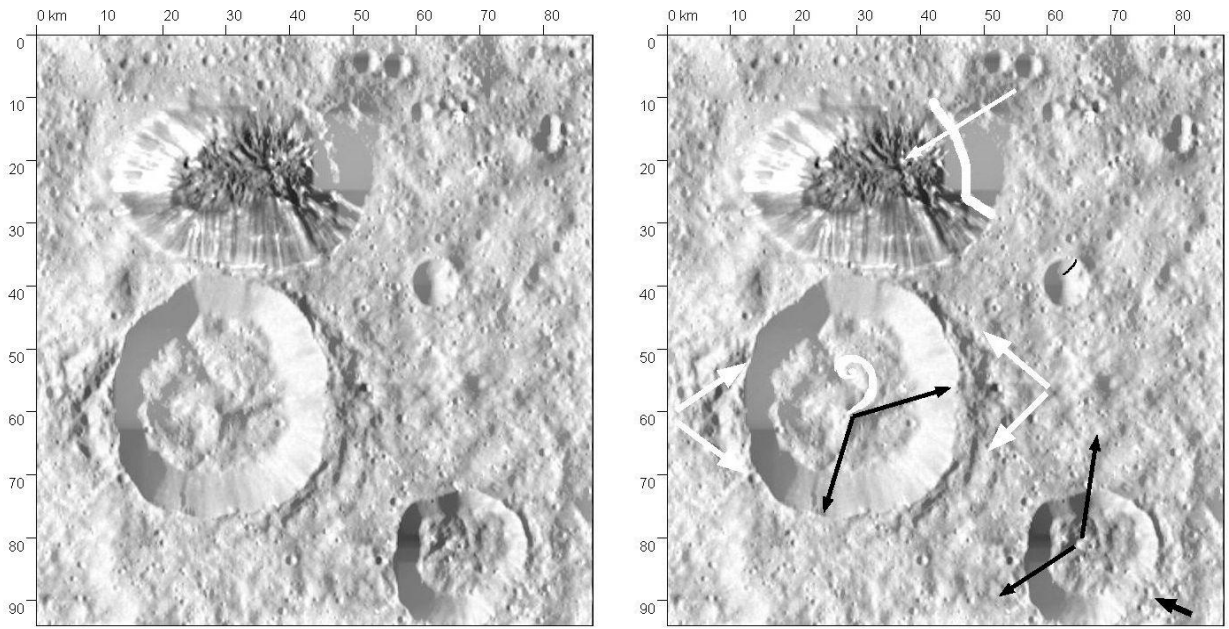

*Fig.11. Crop from fig.2 after digital background subtraction (50 pixels ball rolling algorithm). Note diffuse rectangle centered (49, 20). This is one of producer's contrast corrections revealed by the computer processing of the original image. Right image bears additional marks: Short white arrows show directions of local crustal disjunction net. They are the same all over the image. Pay attention to kilometer sized rock boulders located in stress concentrating areas near starting points of the short arrows. Long white arrow points to the small white substance block. It is the point of local grooves' coming together. Black arrows show discontinuity traces on the mountain and hill tables. The comparison of black arrows' directions allows see angle of hill rotation. Reverse direction of long white arrow coincides to that of the lower mountain black arrow. White spiral and curved hillock black line mask the details, connected to rotational plastic deformation of the mountain and the hillock. Short black arrow points to landslide location. Wide curved white line inside the diffuse rectangle marks bright curved concentric scarp on the crater right slope shadowed by the rectangle (cf. fig 10.).*

The wavy relief right down to the hill appears also to result from hill's frictional braking in this direction. The relief wavelength scale seems to look bigger than that in the mountain's vicinity because of less hill inertia dissipation. Avalanche landslide at the 4-5 o'clock position is thought to be ignited by frictional force projection along the hill's lower right slope. Pay notice to the boulders of brighter substance which seemed to survive the fall and to save their integrity. This means they are stronger compared to darker substances around.

The analysis led to conclusion that the pyramid shaped mountain was thrown rotated out of the layered planetary crust. In the flight it flipped and then landed. The starting rotation was anticlockwise as seen from above the planetary surface. At the landing moment the mountain rotation, now clockwise due to the overturn, and its movement along the surface were damped by friction. The hill probably originated as a result of rotational landslide along crustal discontinuity surface inside the ejected predecessor mountain.

## 3.3 Hill and hillock formation alternatives

At the landing moment upper mountain part appeared to cleave approx. horizontally, separated, and slid due to inertia thus forming mountain table. Still rotating a newly appeared hill landed outside the mountain. Alternative but, to our opinion, less probable option is that the hill was ejected out of the crater later than the mountain. The comparison of smoothed perimeters of the mountain table and the hill is shown in fig.12 (cf.fig.9) Due to smaller mass the hill's rotation angle before its landing was twice larger than that of the bigger mountain.

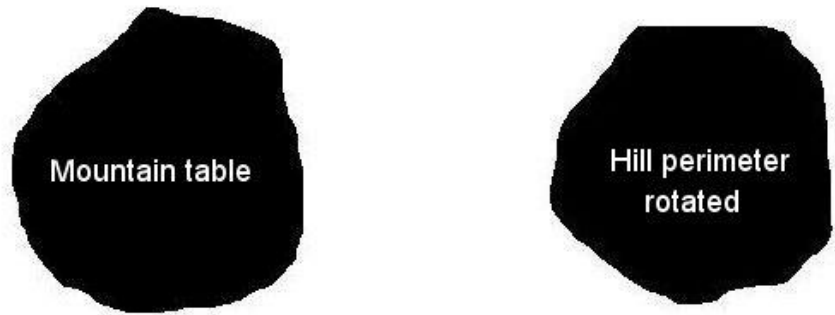

*Fig.12. Areas of the mountain table and the hill cropped from fig.2. The hill area is rotated approx.180° relative to the figure frames. Its upper border is straight because the hill image was cut by the lower frame of fig.2.*

To acquire the impression of the overall hill's shape the low resolution views of Ahuna region are presented in fig.13. Notice that the brightest region of hill's flanks coincides to the brightest one of crater slopes (fig.13 arrows).

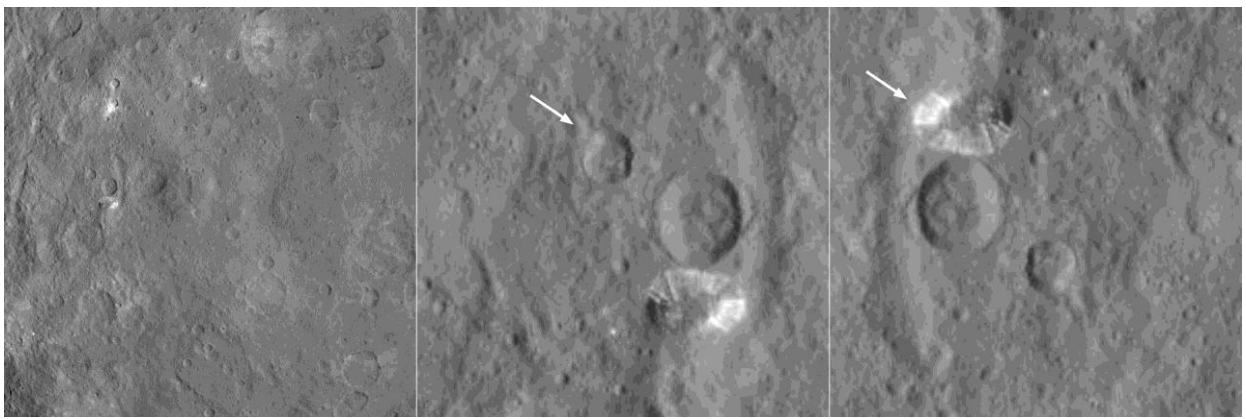

*Fig.13. Left: image is original pia19622 view of Ahuna. Middle image is cropped from the left one. Right image is the middle one 180° rotated to simplify the comparison of relief features. Arrows mark coincident brighter regions. (www.photojournal.jpl.nasa.gov./catalog/pia19622)*

The level of the mountain table was determined by stress concentrating inner crustal border. The stress at the moment of hill's formation exceeded the yield and torsional energy partly released due to plastic deformation. The table relief bears the remnants of the process. One is a spiral or comma shaped feature that is the genetic mark of rotation (see white spiral in fig.11).

Detailed pia20348 image (fig.14 left) delivered by NASA in spring 2016 stimulates our search for the hillock's place of origin.

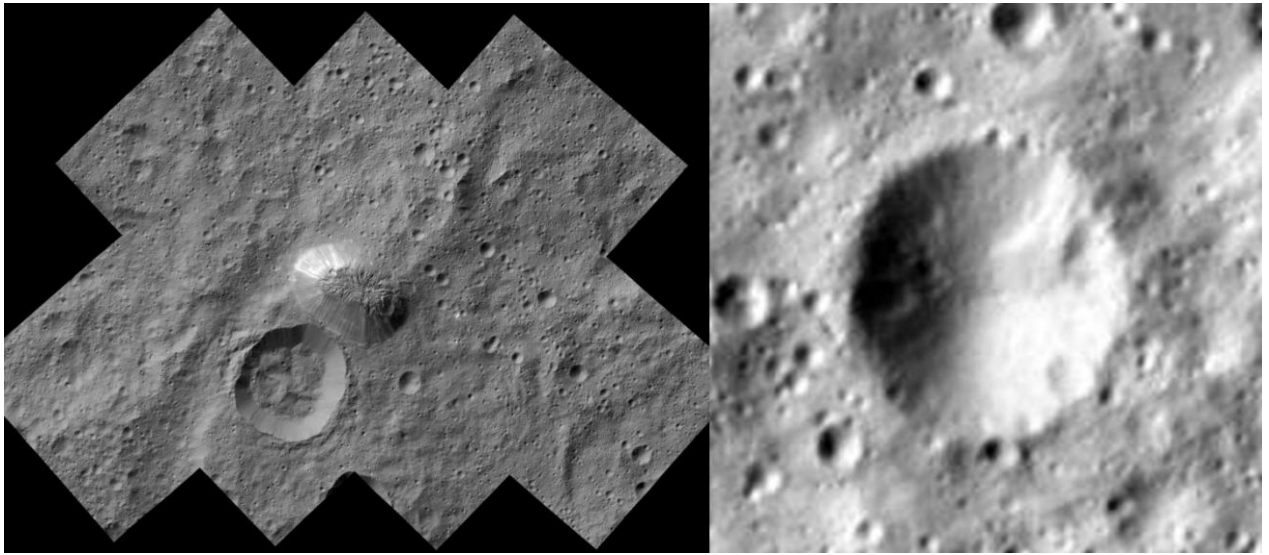

*Fig.14. Left: image of Ahuna region acquired by Dawn from the low altitude mapping orbit 385km above the surface of Ceres. Right image shows rotated hillock's crop from the left image. Its background was subtracted and contrast was enhanced. The orientation of the hillock is the same as in fig.2. (http://www.jpl.nasa.gov/spaceimages/details.php?id=PIA20348)*

Hillock's counterclockwise eject rotation is proved by the curved trace on its slope (figs.11,14). We looked for its ejection position in the region pointed by the long white arrow in fig.11. Hillock shapes appear to coincide to the area inside the crater near to the branching fracture (encircled in fig.15). Whether a summit separated from the hill to become the hillock the same manner the hill cleaved from the mountain, thus way finishing the process of sequential slide, or the hillock was ejected later somewhere in the encircled region, stays unclear. Moreover, the lack of pronounced streaks all over hillock's surface leaves doubts about our correct choice of its birthplace.

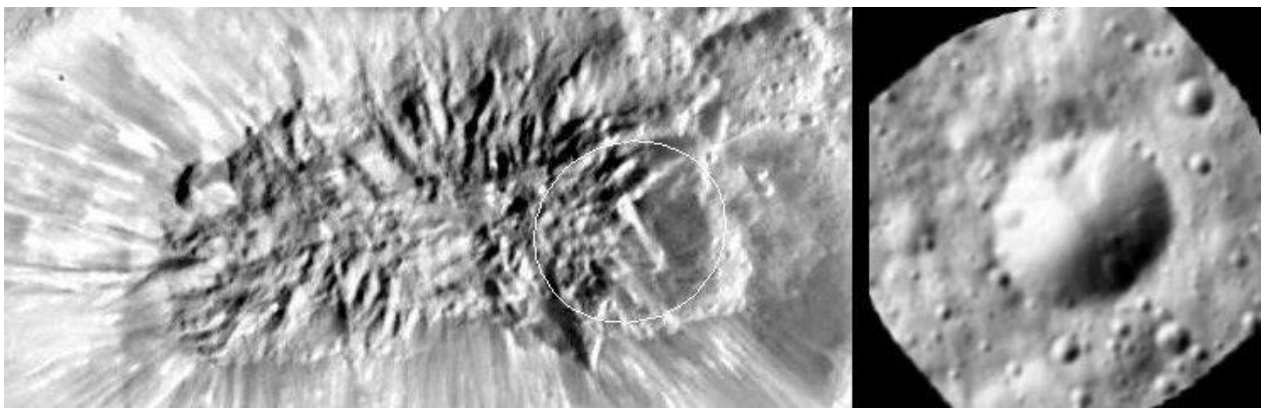

*Fig.15. Left image is crater bottom region cut from NASA's pia20348 image shown in fig.14. The place of possible hillock origin is restricted by white ellipse. Right image is the hillock region cut from the same PIA20348 view. It was flipped horizontally and rotated approx. 45 degrees clockwise (cf.fig.11). The backgrounds of both images were subtracted and their contrasts were enhanced.*

There are also many alike but smaller hillocks in surrounding areas (e.g. see fig.14). The hillock seems to continue straight ca. vertical line dotted by other hillocks of smaller sizes to the right of the crater and the mountain (see fig.2). The line is parallel to the crater's minor axis.

Besides, the hillock is located inside approx. twice larger rhomb-like crustal pit (see figs.14,15(right)). Approximately the same rhomb-side directions are shown by short white arrows in fig.11. So the hillock could be brought about in place of its location due to the same process of ejective orogenesis, but with specific features discussed further on example of Kupalo region. Thus the hillock's way of birth needs further scrutiny and additional observational data.

We described some phenomenological proofs of the crater formation with simultaneous explosive ejection of the mountain. There exist lots of others, both direct and indirect. For instance, the approximate equality of crater volume to that of the mountain and the hill, the equivalence of the objects' slope inclinations, the position of crater's ejecta blanket elevation opposite to the mountain, etc.

We refrain ourselves from consideration of them. Quite enough evidences to support the existence of a thus far unknown phenomenon of explosive ejective orogenesis with simultaneous crater formation were given. Further we study its specific features and consistency with other observational data on Ceres.

## 3.4 Kupalo regional relief formation

Another interesting example of ejective orogenesis is southern hemisphere 26km wide Kupalo crater with very bright slopes (fig.16).

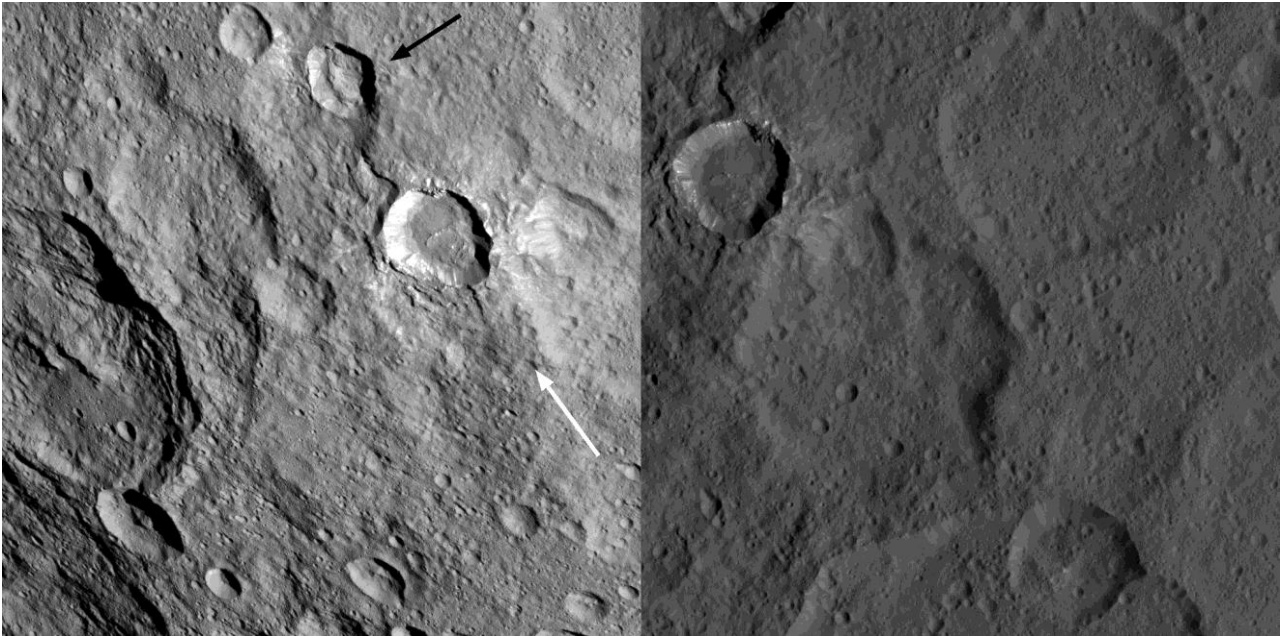

*Fig.16. Images of Kupalo region. Right one is original pia19982, left one is pia19981 enhanced in contrast. White arrow gives crustal fracture direction, black arrow points to plausible mountain cap (see text).*
*(www.photojournal.jpl.nasa.gov./catalog/pia19981*
*www.photojournal.jpl.nasa.gov./catalog/pia19982)*

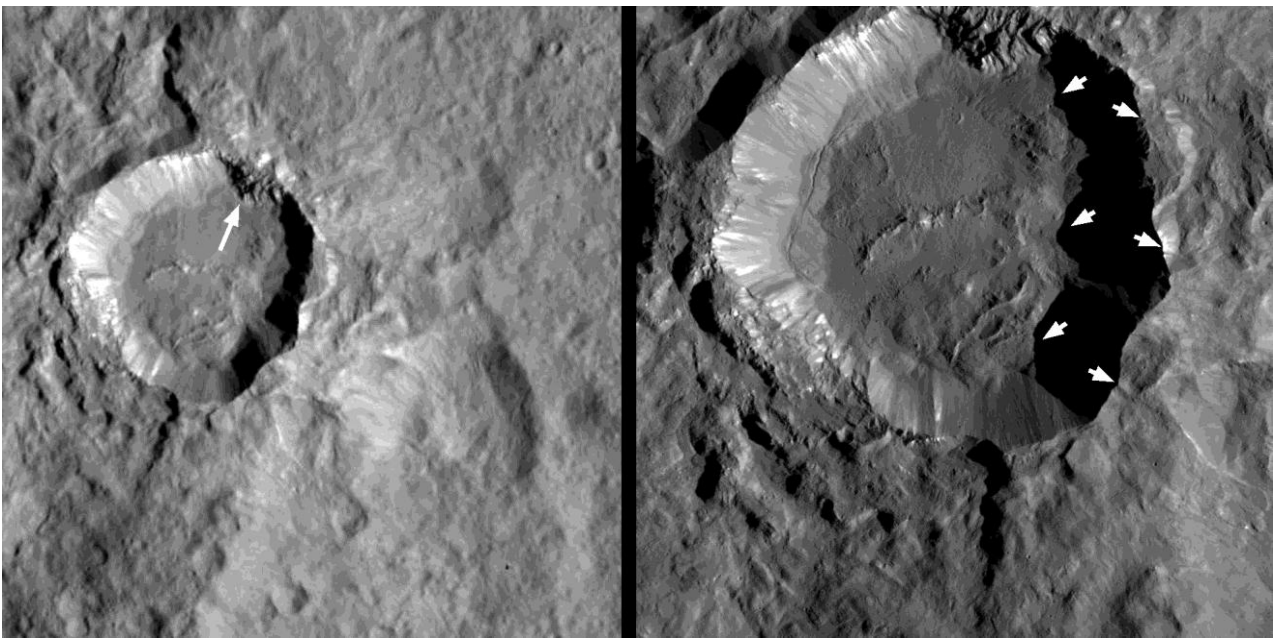

*Fig.17. Left image is a contrasted crop from fig.16(right). White arrow points to a landslide of specific shape. Right image is pia20192 view. Down-left arrows nail the table border of the mountain, down-right - the shadow terminator line.*
*(www.photojournal.jpl.nasa.gov./catalog/pia20192)*

First of all we will prove that geologically fresh Kupalo is a table mountain. Remember, shadows are the keys. If sunlight falls from down-left and left directions in figs. 16 and 17, than the right hand part of the mountain table perimeter casts shadows to its upper right and right vicinity. Let us compare two border lines marked by two approximately vertical groups of short white arrows in fig.16(right).

Enlargement and color inversion of fig.17 allow conclude that the line marked by down-right arrows is not continuous and is crossed by brighter boulder looking objects on the floor (see also fig.18). Several of them even rise above shadow terminator level. Some are elongated and run even perpendicular to the shadow line. In fig.17 especially interesting is the region near upper down-right arrow (see also fig.18). Bright boulders appear to result from partial mountain perimeter breakage due to landing on uneven surface or later slope downfall destruction because of weathering and crustal expansion.

Now compare three different shadow casts in fig.18. All three images available were shot for slightly different lighting conditions and Dawn's telescope inclinations relative to the planetary surface. So the shadows are casted a bit differently.

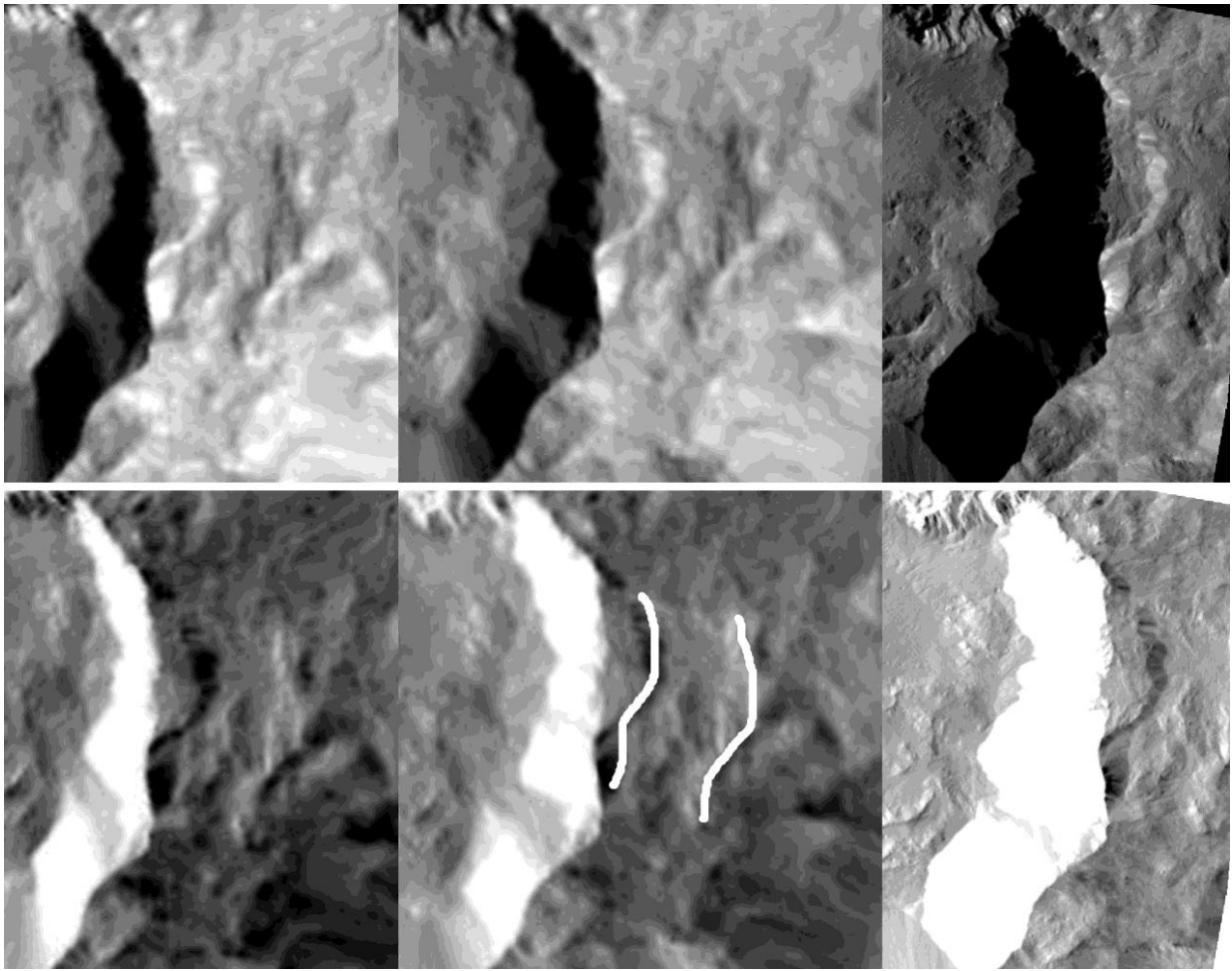

*Fig.18. Shadows in case of a bit different sunlight and shot angle conditions. Upper row, left to right: rotated cuts of pia19981, pia19982, pia20192 views (figs.16,17). Lower row depicts the same views inverted in colors. White crescent-like lines mark a couple of approximately parallel relief features*

It is easy to find contrast difference predetermined by relief features in partly shadowed and slightly inclined terrace areas near the table border edges marked by down-left arrows. This is impossible for shadow line of the floor.

Pay special notice to the brightest region near the middle down-right arrow where different shadow conditions give different terminator shapes (figs.17,18). Here the terminator curvature variations do not coincide to any physical relief variations of the brightest floor region. The terminator line does not cohere to the relief around. Instead it repeats the kinks of the mountain table border. We conclude that it is the mountain table perimeter marked by down-left arrows that casts the shadows.

Okay, Kupalo is the mountain, but where is its crater? The answer is surprising. It is underneath. To prove we compare the relief borders of the mountain and plausible crater rim remnants. Partly degraded crater remnants are clearly seen in all views due to their brightest crescent-like shapes. They are located opposite to the brightest mountain slope across fracture line of approximate symmetry (fig.16).

Topologically similar relief details of crater and mountain slope remnants are shown in fig.19. The likeness of relief features depicted by dark curved lines is obvious.

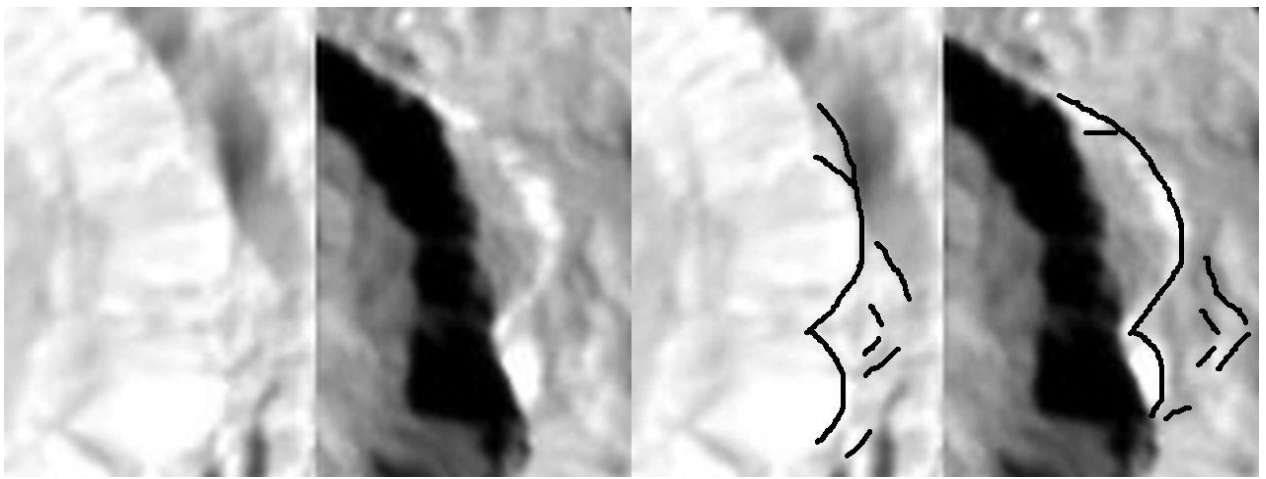

*Fig.19 Crops from pia19982 (fig.16). Left to right: mountain bright slope border and its vicinity flipped and rotated, shadowed mountain slope and its vicinity with crater crescent remnants rotated, the same crops with similar relief features shown by black lines.*

Now we pay our attention to the fracture that runs under the mountain and is the reason of Kupalo's existence (see 5.3). Its direction is shown by white arrow in fig.16. The fracture belongs to regional group of parallel fractures. It appeared to largely widen due to crustal expansion and to form rift valley or partly ruined canyon with parallel borders (see figs.16,17).

We conclude that parallel layer over layer movement of the crust makes crater slopes separate into narrow brighter rim ledges. As we already know they are stronger than darker crustal substance so they are able to form parallel relief features, which save their shapes (figs.18,20). The same crustal tension is exerted on the mountain, strives to elongate it, and partly destructs its borders.

Kupalo mountain is placed above its crater. The major axis of the table mountain is directed along the fracture. The elongated crater under the mountain could result in deviation of its table from planarity. Fig.20(right) shows contrast differences of the mountain table developed after computer contrast enhancement. Averaged contrast gradient is directed approximately perpendicular to the fracture. The result appears to prove our conclusion on uneven mountain table.

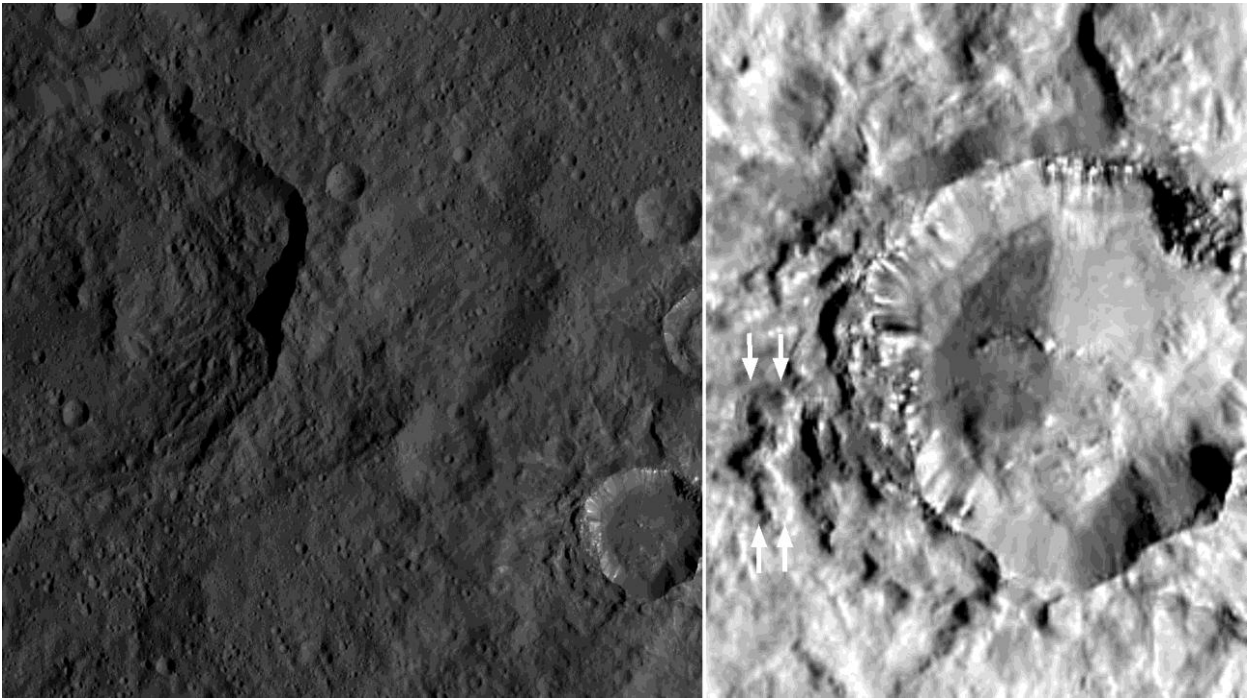

*Fig 20. Left is high altitude image of Kupalo region. Right image is the crop from the left one with subtracted background and enhanced contrast. White arrows point to approximately parallel zigzag relief features.*
*(www.photojournal.jpl.nasa.gov./catalog/pia19903)*

Other interesting Kupalo regional brighter objects are smaller mountains located along the fracture diagonally left up to the frames of fig.16(left). The bigger of them (officially designated as Juling crater) is pointed by a black arrow. It looks like separated Kupalo's cap of elongated asymmetrical rhomb-like shape. The cap seems fresh, bears slope streaks, and appears to have the same lens-like landslide feature as pointed by white arrow in fig.17(left). Landslides of the kind seem to be inherent for explosive orogenesis. The smaller mountain may also be the part of the cap, slid down the nearby elevation (see fig.16(left)).

We do not insist on correctness of above proposed scenario of smaller mountains' ways of birth. They may have originated independently of Kupalo. Alternatively Kupalo's cap may possibly be the elevation to the right of it in fig.16(left). The problem needs further examination, especially spectroscopic.

## 3.5 Kupalo-like objects

There are lots of objects with the formation history like that of Kupalo on Ceres. In fig.21 we present an interesting image showing the ejected mountain with flank relief detail exactly repeated by the slope of underlying crater.

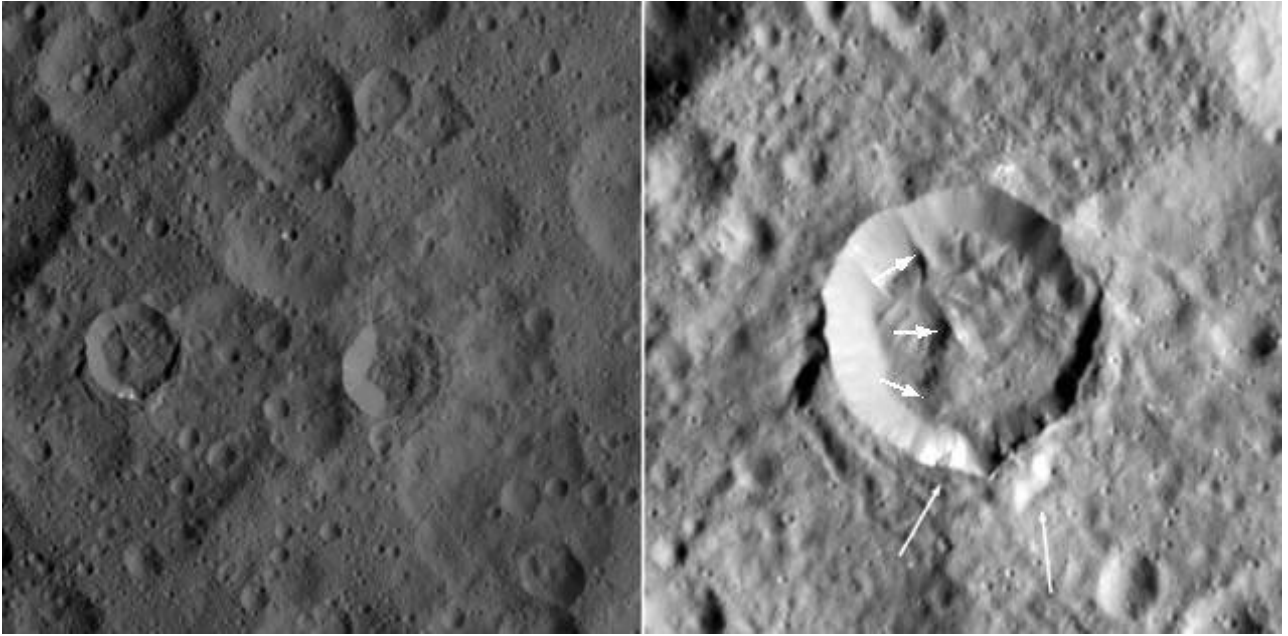

*Fig.21. Left image is original Dawn's pia19997 view. Right one is its contrasted crop with twenty kilometer wide Czosoby object inside. Long white arrows point to two beak-like relief features mirroring each other. Short white arrows nail the curved fracture line inherited by the left part of mountain table. (www.photojournal.jpl.nasa.gov./catalog/pia19997)*

The exact coincidence leaves no doubts on the origin of Czosoby. One can easily compare the objects' rims and their colors. It is interesting to notice that former concave fracture line is clearly discernable at the mountain table (Fig.21 short arrows). The role of expanding fracture or rift valley is played by the round rim border of twice wider elevation to the right of the ejected mountain. This at first glance surprising fact does not contradict to our tensional stress approach to physical mechanism of ejective orogenesis (see 5.3).

The beak-like mountain rim shapes (fig.21 left arrow) seem to be natural for explosive orogenesis. Beak position appears to be always near the fracture line, the reason being local shear stress release. Probably the beak vicinity is the region of maximal white substance accumulation due to the severest plastic reshaping, chemical modifications, and unique detonation (see 5.3, 5.4). For instance, see similar features of Ahuna mountain and concomitant crater shapes (e.g. in figs.6,7).

Other examples of ejected mountains located above their craters are presented in fig.22. Crustal fractures around the mountains are shown by arrows. The larger 15km wide table mountain strongly resembles Kupalo. In its case the remnants of similar relief features of the mountain and the crater are seen very clearly.

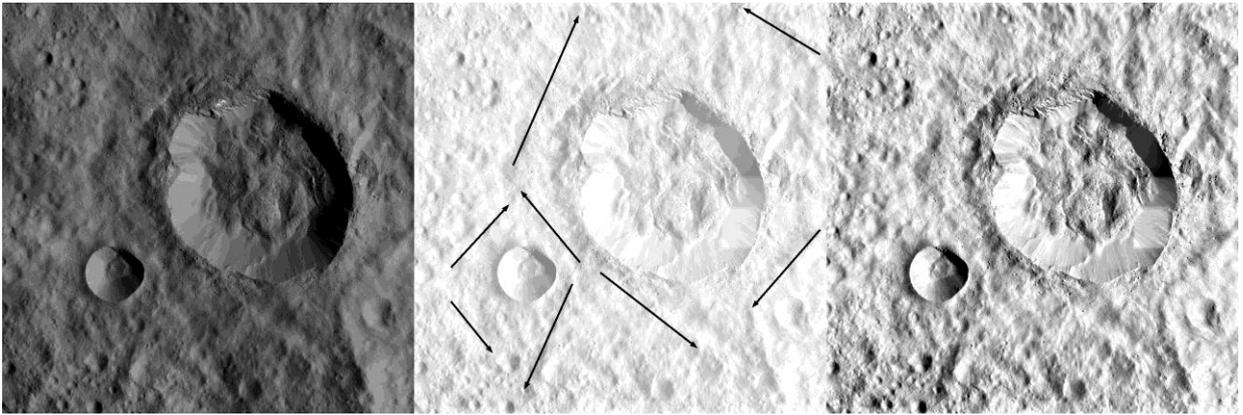

*Fig.22. Two mountains located at approximately 23 degrees south latitude, 279 degrees east longitude. Left photo is original pia20310 image. Central one has its background subtracted. Right one has its contrast enhanced and background subtracted. Black arrows mark the fracture geometry around the mountains. (www.photojournal.jpl.nasa.gov./catalog/pia20310)*

      The smaller 5km wide mountain in fig.22 bears curious table relief features. To explore them we cropped the region of interest out of the original image and magnified it (fig.23). One can see that the mountain table geometry is like that of the fractures shown by dark arrows around the mountain (fig.22), but being of smaller sizes, inverted horizontally, and a bit rotated clockwise (ca.10°).

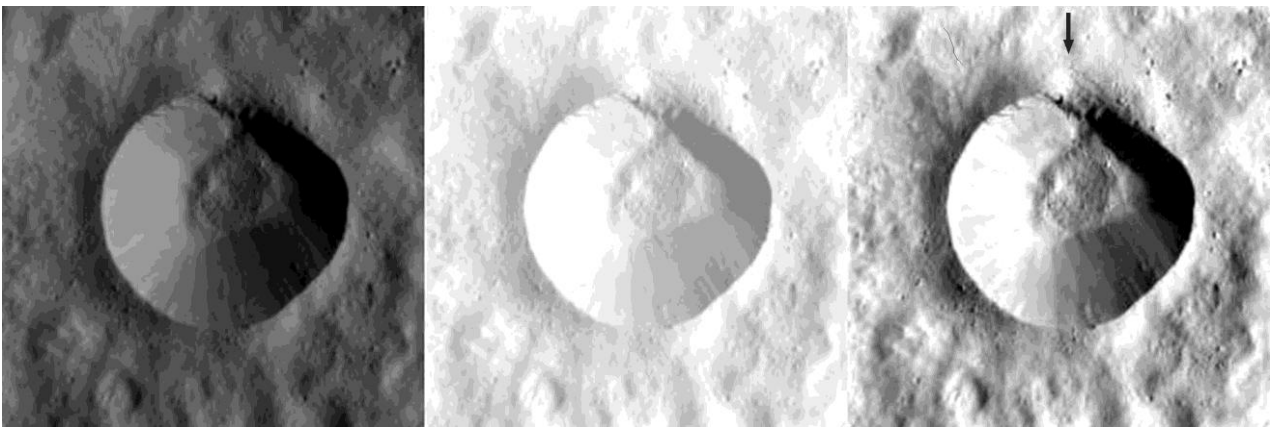

*Fig.23. Crops from fig.22 showing smaller table mountain. Left photo is the original image. Central one has its background subtracted. Right one is contrast enhanced central photo. A black arrow points to a damaged region natural for explosive orogenesis.*

      As a result we infer that fracture symmetry exactly proves mountain's ejection anticlockwise out of the crust and its flip in the flight. We conclude that just before the ejection the future crater area of the crustal surface has been striving to become convex. This is the reason of fractures' planes inclinations to crustal surface and formation of a pyramid by the crack planes. Due to this the buried base of the mountain may not be flat.

The pyramid is truncated by the border of upper crustal layer. Contact stress concentrating surfaces between crustal layers are prone to plastic deformations, material transformations, and fluidization. Two mountains in fig.22 appear to be of the same height because they originated inside the same upper crustal layer. They are so to say the measures of its depth.

Existence of a crater under the cone mountain is also evidenced by light contrast in the vicinity of its perimeter (fig.23). The reason of the contrast is the floor level going down around and under the mountain. So the vicinity is partly weathered crater rim.

It is interesting to compare the smaller cone mountain and the hillock in Ahuna region. Two curved slope lines from hillock's apex evolve to a widening rift valley with bead-like elevation near the right valley border closer to the base (see fig.14(right)). Curved slope line of the small mountain starts from the apex of its table rhomb which is stress concentrating point. The line is the borderline of widening damaged area, may be a landslide (fig.23 black arrow), which seems to be usual for explosive orogenesis.

Small brighter elevation on the crater floor nearby formed as a result of slope destruction. It seems that this damage is the terminal stage of plastically hardened beak-like area destruction. Consequently the considerations of plasticity which leads to local changes of impurity concentrations, formation of new substances, volume and density shifts may explain the shape of the mountain slope features.

Another interesting example of Kupalo-like orogenesis is shown in fig 24. There are several ejected pyramids in the image region. A pyramid located in the upper part of it is enlarged in the figure (right) to simplify the relief analysis the way we applied above.

One can clearly discern the inversion symmetry coincidence between rhomb-like perimeter of ejected pyramid and fracture picture around it. These fractures are diagonal to the figure frames. The local family of parallel discontinuities resulted in regional lined relief elevations to the right and down of the pyramid and to lined hillock laces to the left and up.

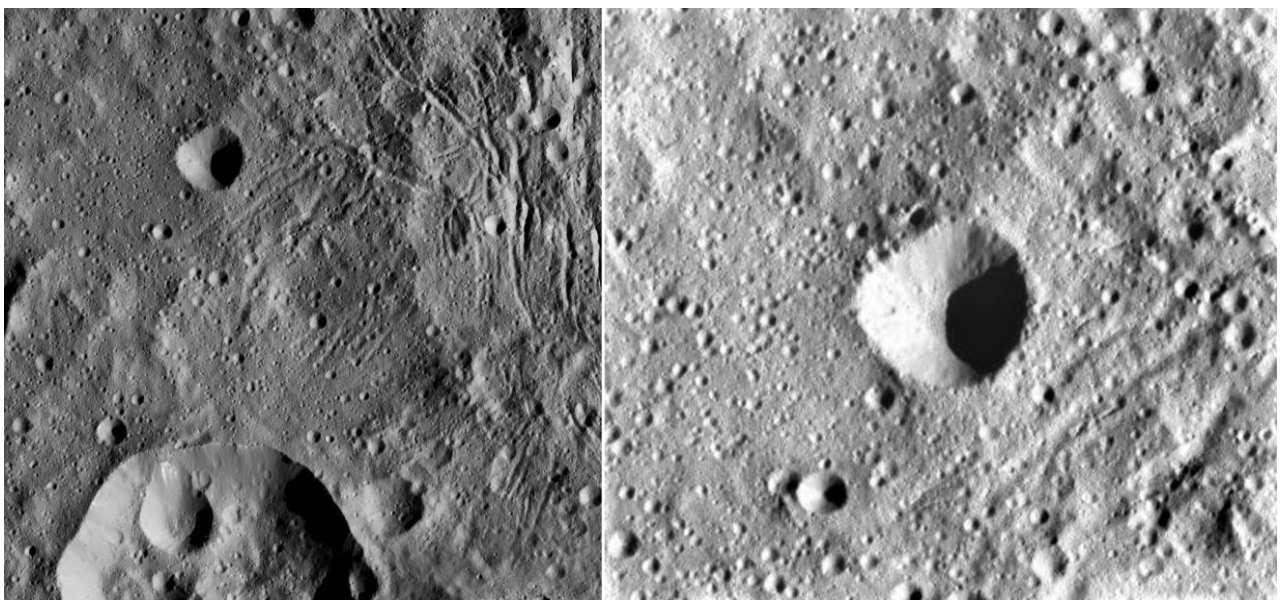

*Fig.24. Original pia20957 image (left) and contrasted crop from it (right). (www.photojournal.jpl.nasa.gov./catalog/pia20957)*

The brightness coincidence between the crater and pyramid rims is also obvious and proves the symmetrical conclusions. The coherence points are easy to determine comparing locations of brighter substances and perimeter curvatures. The analogy to the cone shown in fig.23 is almost total.

The overall relief is more pronounced to the right that is opposite to the direction of the smallest pyramid base angle. To our opinion the direction of pyramid's flip in the course of explosive ejective orogenesis is connected to the one of relief elaboration which determines the detonation (see 5.4)

The outcome of the above analyses is that crustal fractures and cracks directly determine conical shapes of ejected mountains. In general, overall geometry of ejected objects and their craters is heritably connected to stress geometry and crustal discontinuities' map in places of their formation. The approach is a key to explanation of unusual and highly symmetrical relief features of Ceres as well as other celestial rigid bodies. By the way, this symmetry brought about numerous layman assumptions about artificial origin of improbably symmetrical relief features.

On Earth geologic discontinuities concentrate stress and as a rule alleviate fluid movement from deep insides. Fractures, cracks, and faults are prone to be filled with different impurities, inclusive substances, and even magma. The same phenomenon universal for rock bodies of Solar system holds for Ceres. For instance, this is directly proved by the Cerean planetary maps with stripy distributions of phyllosilicate and ammonia abundances [5].

Horizontal discontinuities e.g. borders between crustal layers play the same role as well as surfaces of plutonic bodies of different sizes and crustal inclusions of various kinds. It is well known in geology that these stress concentrators are also able to ease fluid migration, substance accumulation, and crack formation. Generally, all border surfaces between substances of different composition are stress concentrators which alleviate fluid and impurity migrations on different scales.

## 3.6 Chained elevations

Lined chained elevations as in fig.24 are natural for Ceres. In figs. 25,26 self-organized lined elevations (we prove that Occator is a table mountain in 4.1) of diverse sizes in Occator region are shown. They prefer to orient in directions which are diagonal to the planetary equator. Chained objects on horizontal surfaces appear to evolve due to plastic materials' modifications the same way as just discussed those on slopes. They seem to be connected to interplay of tensions and shears as well as the nature of crustal substances.

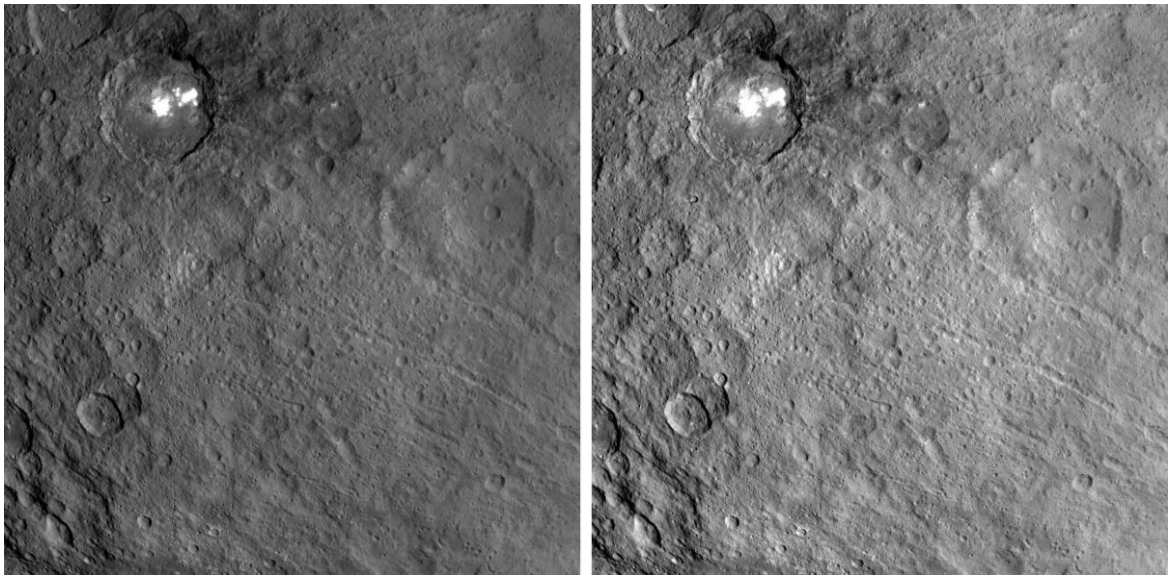

*Fig.25 Right: original image of Occator (left up) and Kirnis (right) regions. Left: the same image with enhanced contrast.*
*(www.photojournal.jpl.nasa.gov./catalog/pia21409)*

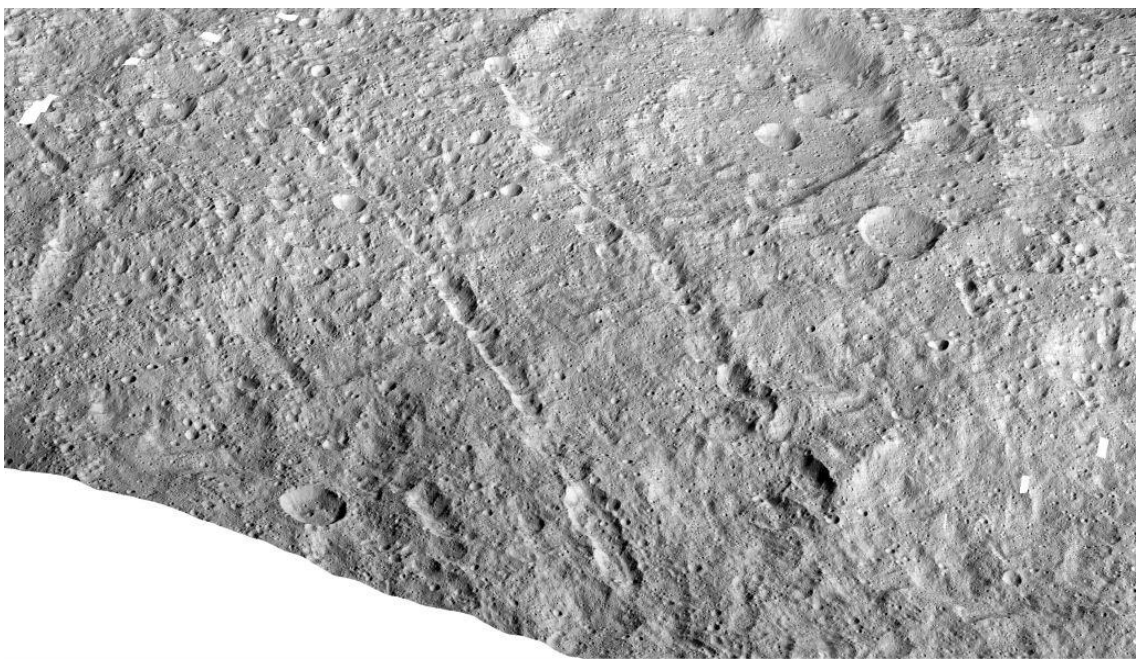

*Fig.26.Image of the area shown in lower right corner of fig.25. The image width is approx. 350 km. This region of lined fractures is called Samhain catenae.*
*(www.photojournal.jpl.nasa.gov./catalog/pia22086)*

**3.7 Planetary fracture system**

Topographic map of Ceres shows some interesting global features (figs.27-31). The most protruding i.e. topographically highest cerean region was named Hanami planum (centered 15ºN, 230ºE). It is the oval area encircling Nawish-Occator-Urvara-Meanderi quadrangle, which diagonals are crossed in Azacca vicinity near equator of Ceres (fig.27). The major axis of the oval is Occator-Meanderi diagonal that angles zero latitude at 45°. The region is clearly highlighted by the concentration of red color in the figure.

Antipodal to Hanami region is also averagely protruding but subtler one (relative to zero level of planetary globe). Regions centered near 130° and 310° longitudes are depressions. So the first approximation of equator section is an ellipse. All four mentioned regions have marks of shear crustal stresses. We showed some of those global rhombs and lineaments with black lines (fig.27).

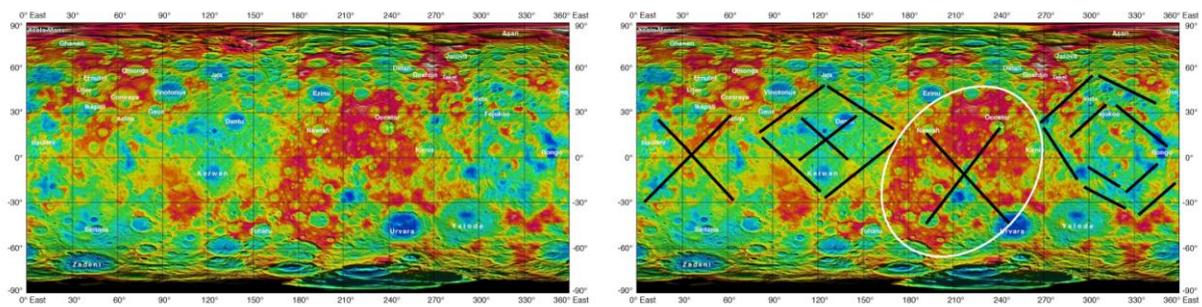

*Fig.27. Right image is the original NASA's map of Ceres. Left image is the same map marked with black arrows to depict some relief features connected to global shear and tensile stresses. White oval shows the position of the largest planetary elevation.*
*(www.photojournal.jpl.nasa.gov./catalog/pia19974)*

The most interesting to us Occator, Kirnis, Samhain catenae, and Ahuna regions are featured in figs.28, 29, 30.

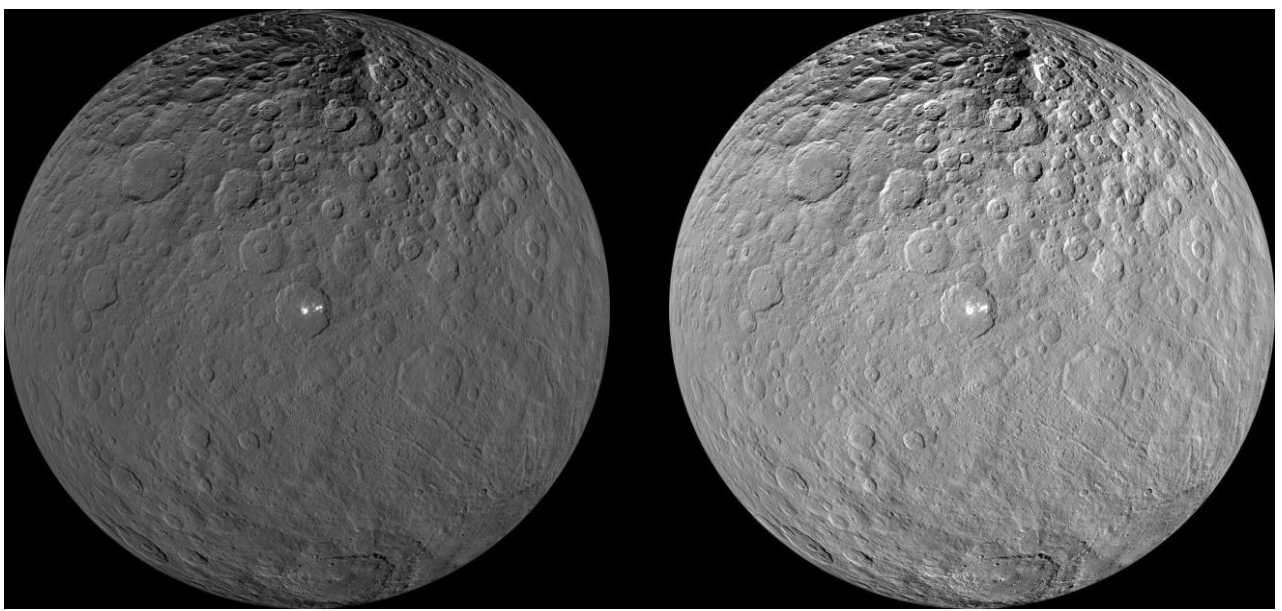

*Fig.28 Original (left) and contrasted (right) images of Occator's side of Ceres.*
*(www.photojournal.jpl.nasa.gov./catalog/pia21906)*

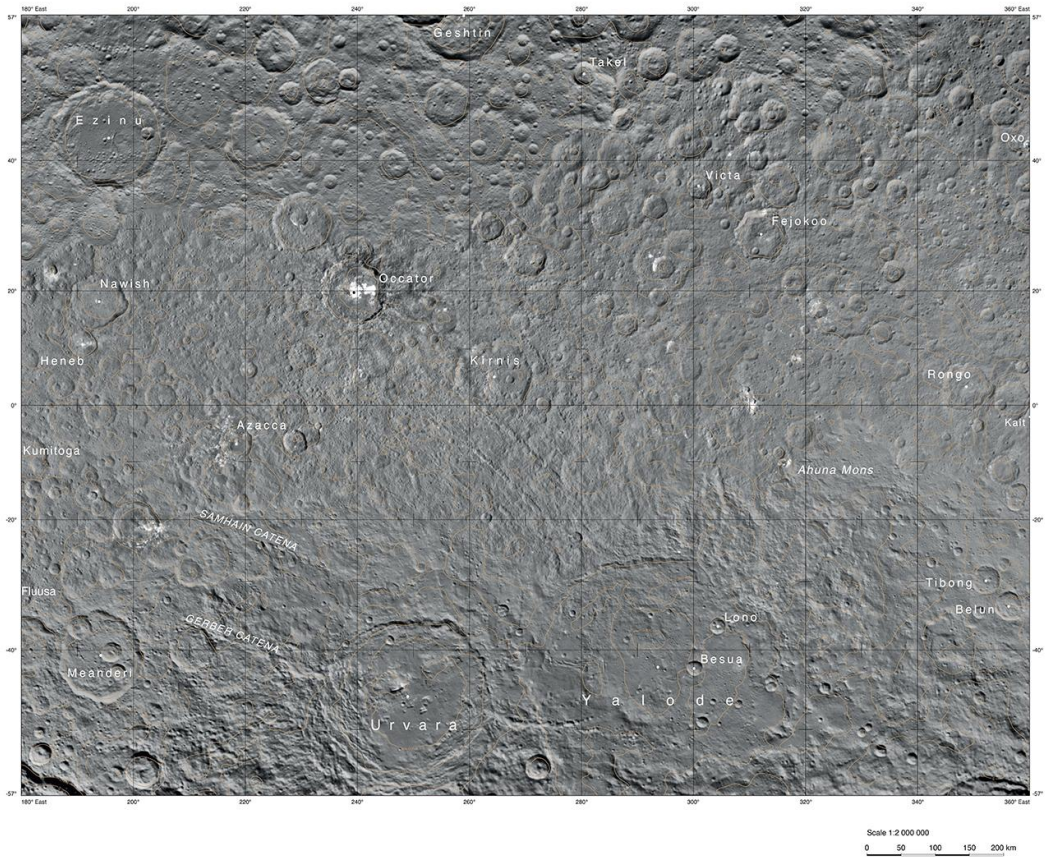

*Fig.29. Occator and Ahuna regional map from NASA' atlas of Ceres. (www.photojournal.jpl.nasa.gov./catalog/pia20014)*

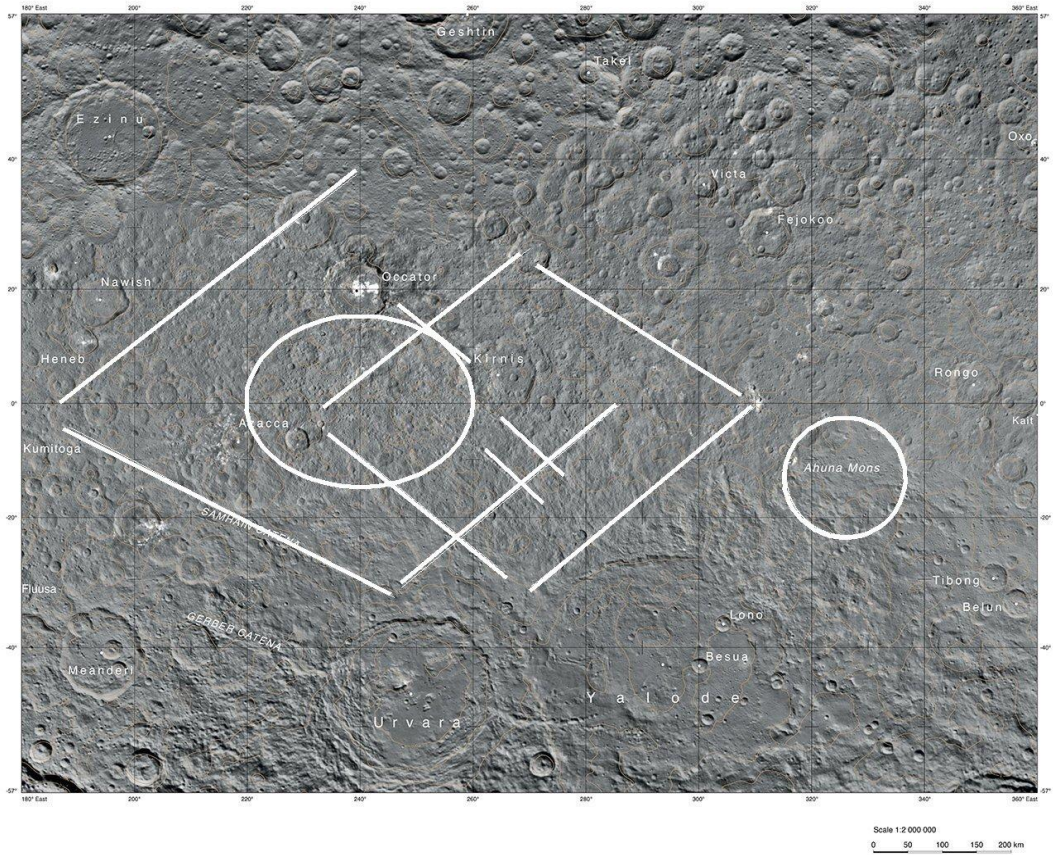

*Fig.30. Fig.29 with white lines showing symmetry and hierarchy of some relief features of Occator and Ahuna regions.*

The examination of the figures leads to conclusion that locations of smaller relief features are connected to (if not determined by) larger ring and rhomb structures. Their positions for Occator and Ahuna regions are shown with white lines in fig.30. The attentive viewer is free to find lots more features of the kind. Such interconnections reflect the hierarchy of planetary mantle and crustal layers. From mathematical standpoint this means nonlinear interaction of planetary modes of corrugation.

Overall relief symmetry proves tensile and shear evolution. We see that remnants of large and global catastrophic events are severely superposed and almost erased from the surface of the expanding planet. Geologic activity of Ceres was tremendously higher long time ago and has been decaying since then.

The area around Occator is of special symmetry. The central large object is surrounded by a hexagonal lace of smaller ones. They are shown by black arrows in fig.31. The hexagon is somewhat extended in latitude direction. The symmetry of the picture appears to be explained by shear stress concentrations due to nonzero torque exerted on Occator crustal region by a rotating batholith-like body floating up from inside

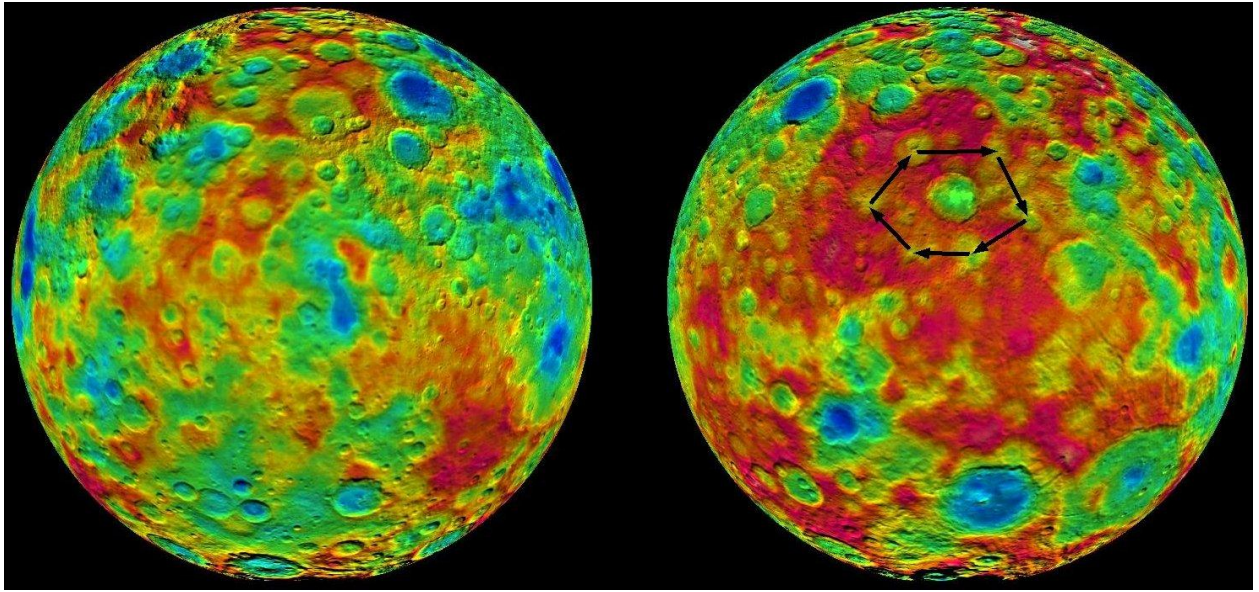

*Fig.31 NASA's topographic map of eastern and western hemispheres of Ceres. Right is marked layered by arrows to show hexagonal symmetry of some relief features connected to rotational stress.*
*(www.photojournal.jpl.nasa.gov./catalog/pia19607)*

Thus we see two kinds of global crustal structures reflected in surface relief, linear and round shaped. The former as on Earth are preferably going along diagonals to the planetary equator (North-East, North-West and opposite). The largest of the latter as on Earth forms global imaginary cube symmetry with two of its facets at North and South poles (polar radius of Ceres is less than equatorial one). The smaller of the latter are at first glance located more randomly. Like lineaments they are sometimes marked not only by relief features, but also by the concentrations of particular substances and minerals. The example of the kind is shown in fig 32. The artificial color palette represents composition and texture differences in regolith materials on the surface of Ceres.

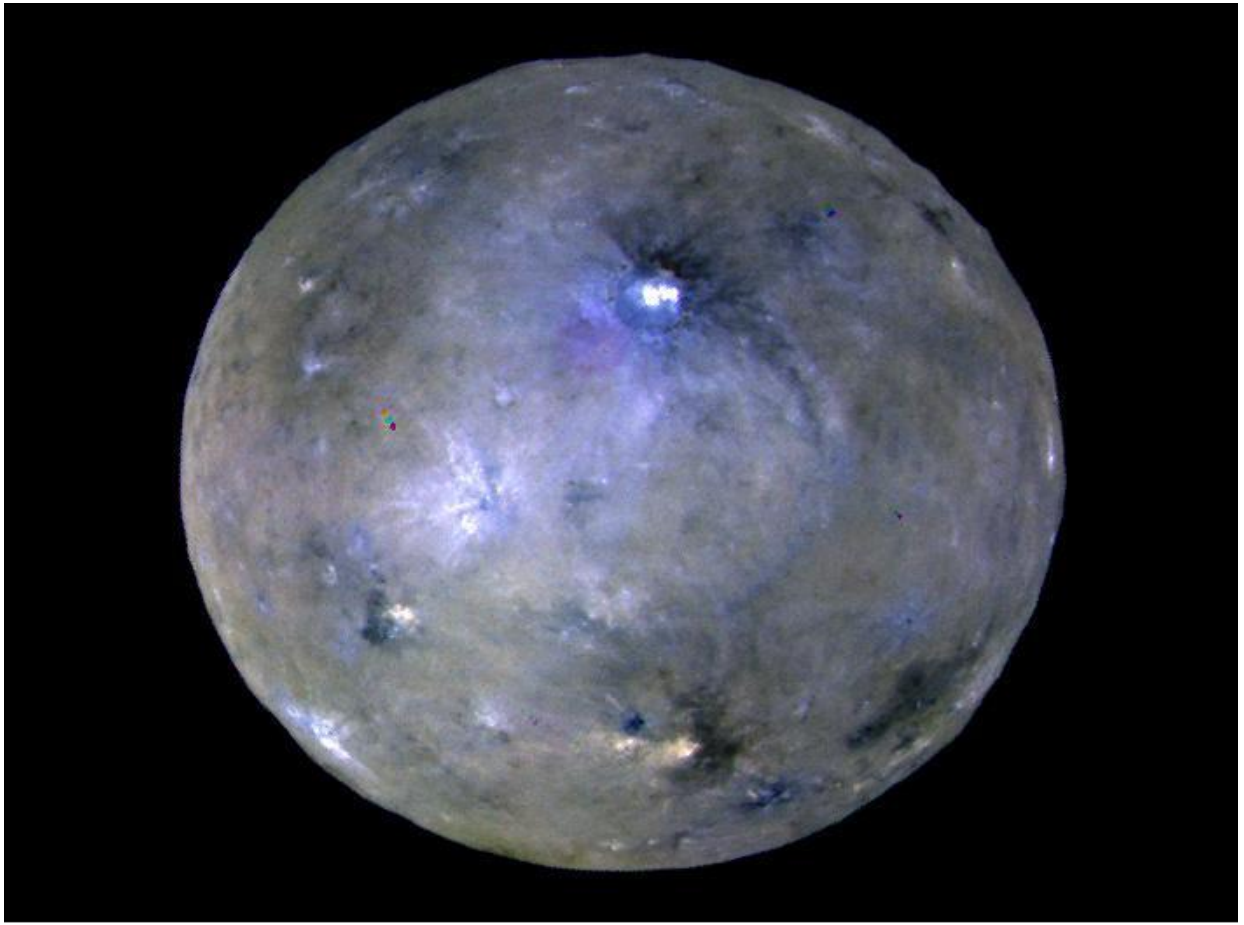

*Fig.32 Enhanced color image of Ceres. The mix of filtered views allows see subtle differences in the composition and texture of the planetary surface.*
*(https://dawn.jpl.nasa.gov/multimedia/images/image-detail.html?id =PIA21406)*

According to the original image description the stunning violet circa concentric to cerean perimeter circle, materials of which differ from the ambient (fig.32), is the result of extraordinary ejecta redistribution and deposition onto the planetary surface. The circle crosses Occator, which is thought of as formed due to outer impact. The concentration of violet color is higher around it, but mainly to the North-East of it.

Contrary, we think that the circle, that restricts part of Hanami planum, reflects the inner structure of cerean mantle and crust. There is the possibility of some of inner materials moving upward. The negative residual gravity anomaly in Hanami planum [6] indirectly confirms the conclusion. To our opinion circular perimeter is marked by inclusions alien to the bulk of local crustal substances.

## 3.8 Landslides and cold cohesion

In this section we give examples of sliding process. It is quite usual for Ceres as well as for other rigid celestial bodies. Discussed earlier hill's formation in Ahuna region is one of its examples. Several other views of Ceres with sled objects are shown in figs.33,34. The places of possible relief coincidence are marked by white and black lines in fig.33. One can see that sliding is responsible for topological similarity of the chosen relief objects.

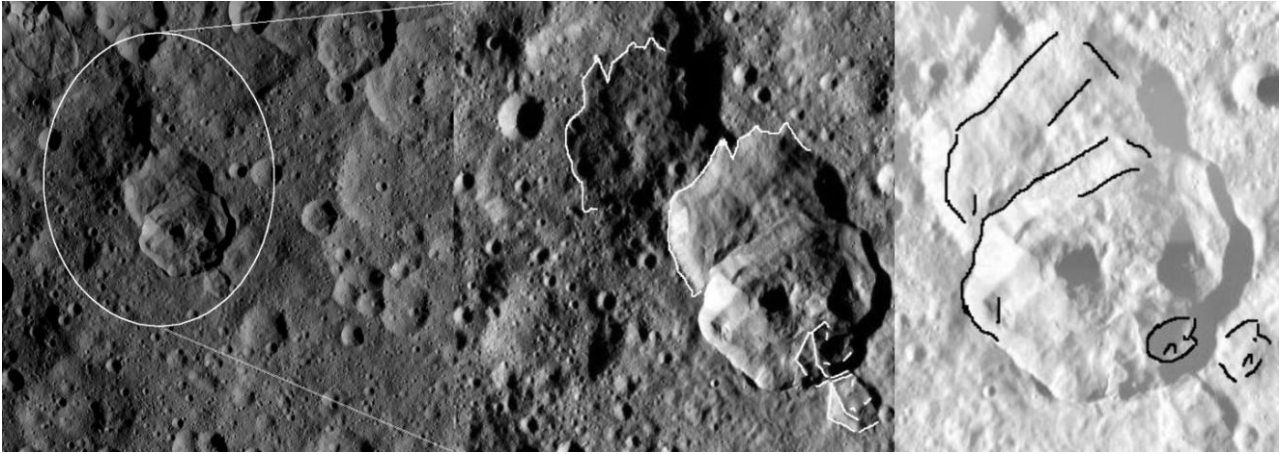

*Fig.33. Dawn's pia19635 image. Left one is the original view. Middle image is a crop from the left one after background subtraction. Right image is the contrasted crop from the left one. Some places of topological coincidence are shown with white and black curves.*
*(www.photojournal.jpl.nasa.gov./catalog/pia19635)*

In fig.34 the similarity of some large relief features is clearly seen without marking curves.

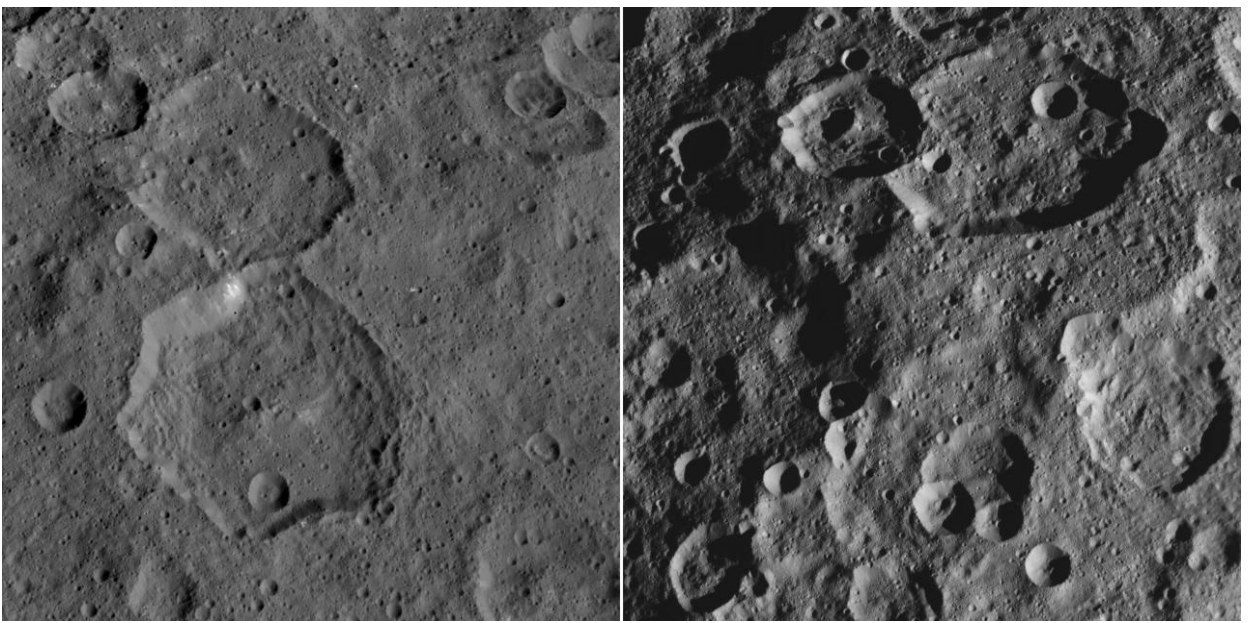

*Fig.34. Two original Dawn's images showing several sled objects.*
*(www.photojournal.jpl.nasa.gov./catalog/pia19908*
*www.photojournal.jpl.nasa.gov./catalog/pia19971)*

So the conclusion is that sequential sliding strongly influences planetary relief of Ceres. Whether it is explosive or slow in time the result is the formation of fractal surface relief. Underlying physical and chemical processes appear to have much in common for both sliding and ejective orogenesis. But sliding is less powerful process than ejective orogenesis for objects of equal sizes. That is why it should prevail in cases of calm planetary geology.

We believe that in cases of sliding alleviation, frictional decay, and boulder movements unexplainable by simple gravity the same approach as for ejective orogenesis should be applied. Velocity enhancement is to be mediated by plastic deformation and sometimes formation of flowing and chemically modified liquid or even gas like layer under and inside the sliding object. Those are final results of shear and tensile stress initiated phase transitions.

In Earth's conditions the generation of water and gas (e.g. methane) during sliding is well known to oil geologists and mine engineers. Sturzstroms in Alp regions are famous for their unusually long pathways unexplainable by dry friction and gravity forces only.

After formation of large objects the process of ejective orogenesis continues on smaller scales. The same holds for sliding. The reason is that ejected or slid objects after some time closely contact planetary crust and cohere to it because of diffusion. Thus little by little they acquire all crustal stresses and become parts of crustal upper layer.

But borders of mountain-crust cohesion are still stress concentrators and fluid conductors. So after some time in case of growing stresses all the processes are possible to be repeated. They will easily go in the places already stressed, for instance, around plastically deformed inclusions or near disjunctions of different types.

According to Dawn's data almost one third of cerean craters have slides near them. The fact is a conundrum for observers. From our point of view a great number of them are mountains in fact. Therefore, in the logic of discussed ideas the observation is quite natural.

# 4 White spots of Occator

## 4.1. Central spot of Occator

We see white substances in all Dawn's views of Ceres, but inside Occator they are rich and bright pile-ups located in two areas (fig.35). The main one, Cerealia facula is in the Occator's center. Other, less bright Vinalia faculae situated to the East of the center in the local fractures' direction (fig.36).

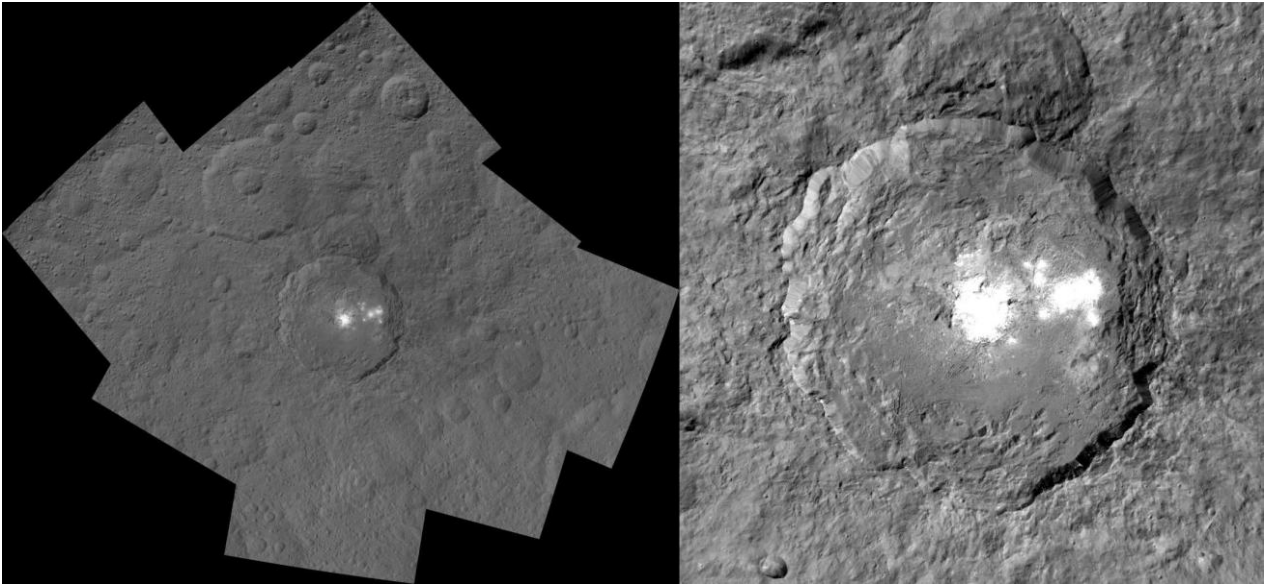

*Fig.35. Left: Occator crater image with famous white spots inside (North is up) It was acquired from the orbit 1470km above the surface. The original view is a mix of two photos for contrast improvement. Right: digitally contrasted crop from the original image to make spots clearer.*
*(www.photojournal.jpl.nasa.gov./catalog/pia19996)*

Occator's rim is evidently separated from vicinities, despite the fact that the eastern and southern borders are widened in a staircase manner. For a crater it looks suspicious.

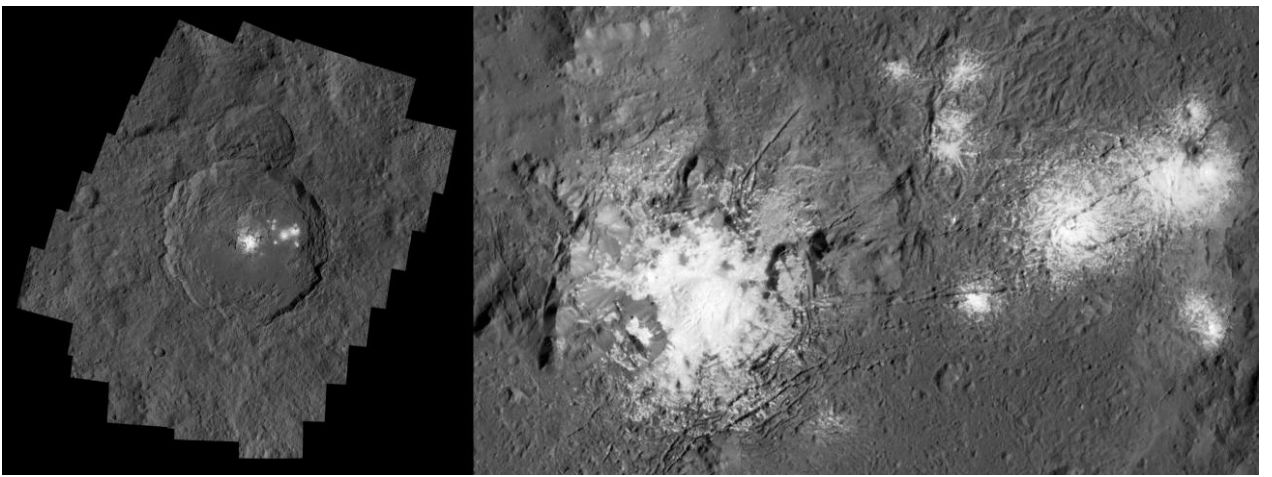

*Fig.36. Left: One of Occator's images. Right: contrasted crop from the left image.*
*(www.photojournal.jpl.nasa.gov./catalog/pia20350)*

Contrary if Occator is a table mountain, strong undestroyed cliff toes above weak regolith layer are to be natural and result in clear contrast with surroundings. We consider Occator as a table mountain and shall prove it in the course of our study.

In fig.37 the central part of Occator looks like usual volcanic edifice. The brightest detail of fig.37(right) is approximately vertically elongated (relative to the figure's frames) object with an oval head exactly in the center. Its length is several times more compared to its width. It reminds Earth's dyke or enlarged and shortened version of a cerean crater slope streak. It is posed along the bunch of crustal fractures. Following the logic of our approach we conclude that the fractures determined the shape and position of this dyke which ripped up the crust.

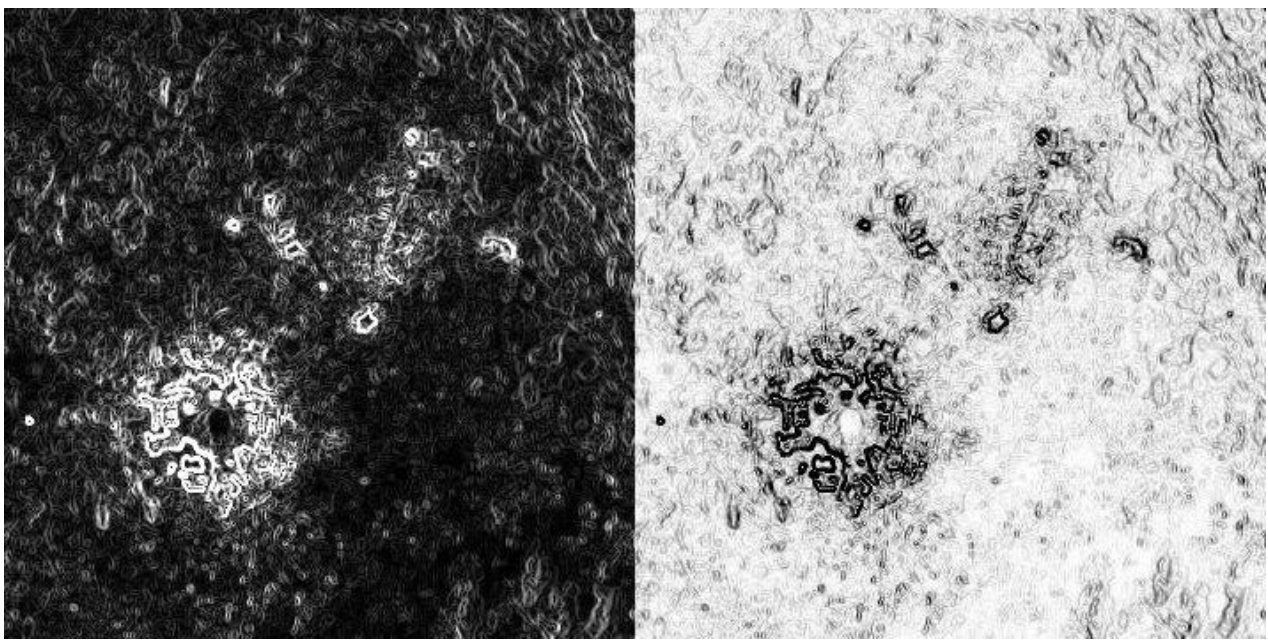

*Fig.37. Left: Crop from pia19889 after visualization of borders of maximal contrast (edge detection algorithm). Right: inverted left image.*
*(www.photojournal.jpl.nasa.gov./catalog/pia19889)*

High resolution image made from low altitude mapping orbit allows us to see details of the region, especially its central part (fig.38). Strong relief contrast lines there are prone to be radial and concentric. The central white depression is encircled by lots of fractures, especially South-Eastern part of it. Destruction in approx. southern direction is less severe, compared to the opposite one, seemingly due to the line of dyke's formation and thickening

The whole appearance of Cerealia facula is of nothing special. It is quite natural for volcanic structures on Earth and other rigid bodies of the solar system. But it is rather difficult to decipher the exact history of Occator's volcanic activity. Nevertheless the end stage of the process is for sure connected to simultaneous explosive formation of central caldera and white substances inside and around it.

To discern relief details we cropped the central spot from the original Dawn's view in fig. 36(left). Then we rotated the crop the way our vision to take its center as a pit (fig.38). In this case the angle of sunlight incidence is nonzero and it falls from the upper to the lower frame of the figure.

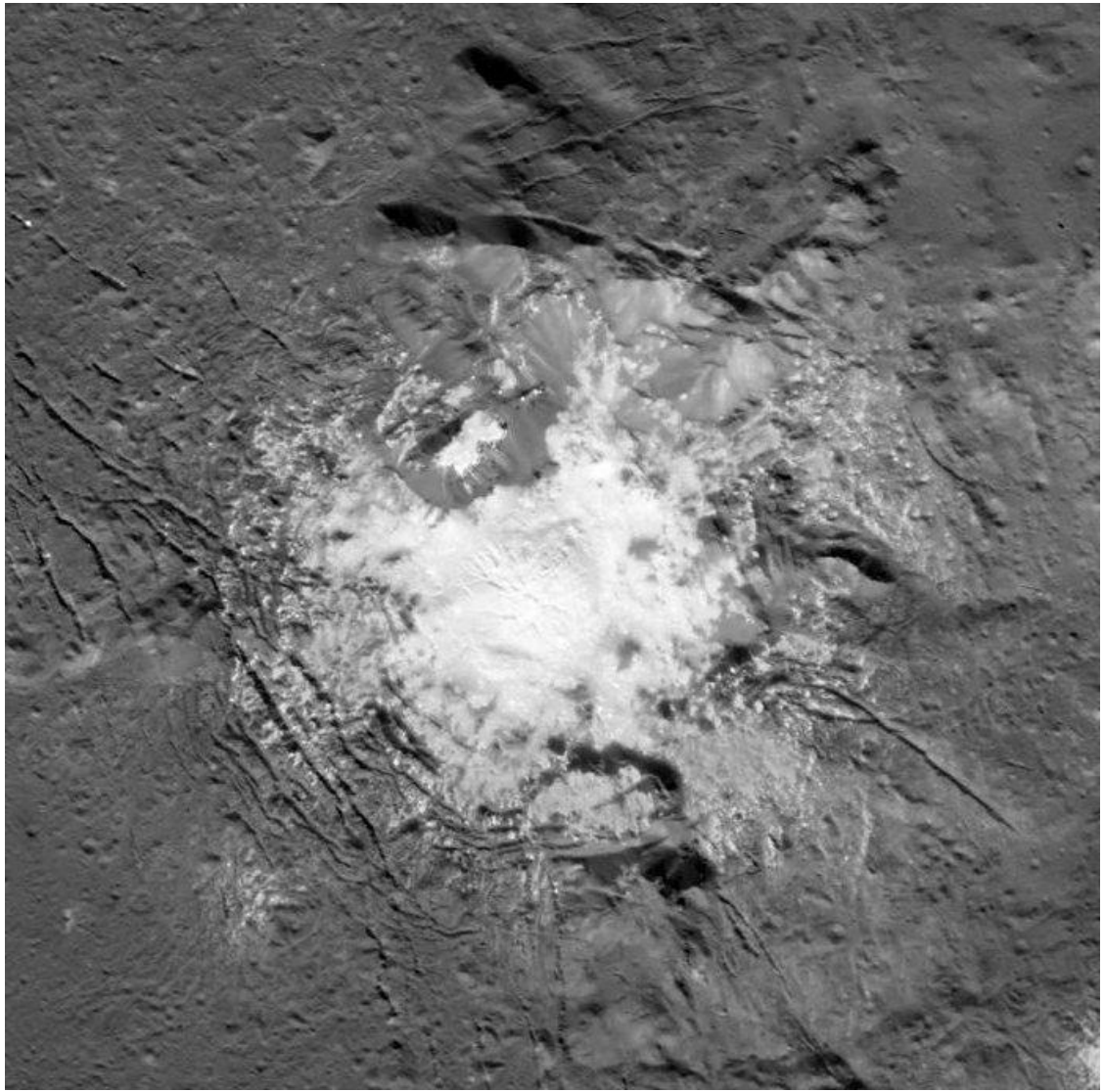

*Fig.38. Enlarged crop from fig.36 (left) features Cerealia facula and Occator's caldera.*
*(www.photojournal.jpl.nasa.gov./catalog /pia20350)*

      Occator's caldera bottom covered with the carpet of white substances looks similar to the floor of Ahuna crater and also has streaks and grooves. The caldera shape is that of plastically deformed zone (see 5.3), its borders give the impression of butterfly wings.

      Now it is easy to find caldera's cap i.e. the mountain thrown out in explosive event of ejective orogenesis and white substance formation. It still saves its integrity. See dark cliff like relief detail up and a bit left to the caldera (fig.38). It is of the same size as the caldera and partly covered with white substance ejecta blanket. We are comparing the features of the two objects and analyze their topological similarities in fig.39.

      Besides, notice that the central caldera's part coincides to the white substance pile-up lying above the overturned cliff mountain. The pile-up has a two-bump appearance and is shown by black arrow in fig.39. To make this result clear we change the colors of the view and marked with black line the averaged border between yellow and red colors (fig.40). So the white pile-up is approximately mirrored in the relief of caldera bottom.

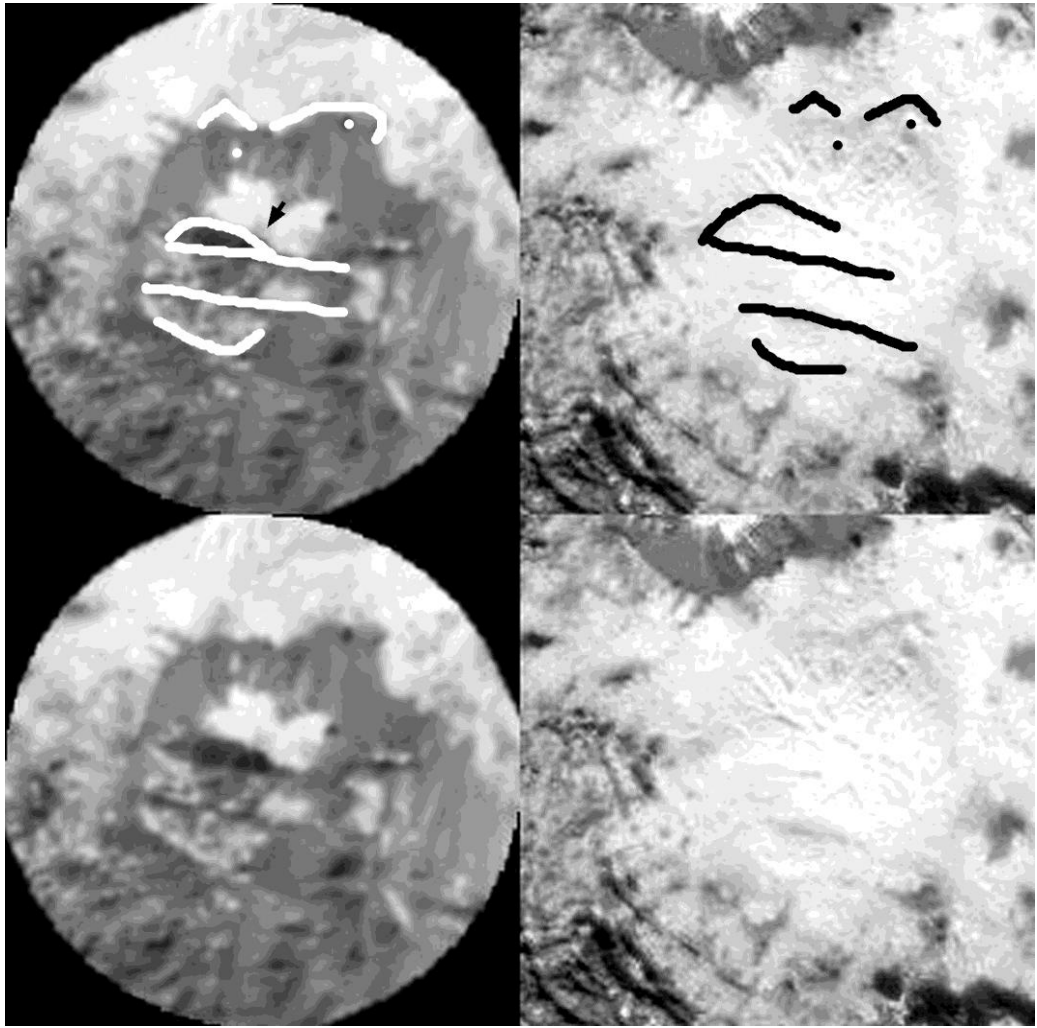

*Fig.39 Left column is a cropped view of cliff-like mountain next up to the caldera of Occator. The view was vertically flipped and rotated clockwise. Right column shows central part of fig.38 with the caldera, covered by white substance. Upper row features coincident relief details. Black arrow points to the approx. middle of white substance pile-up.*

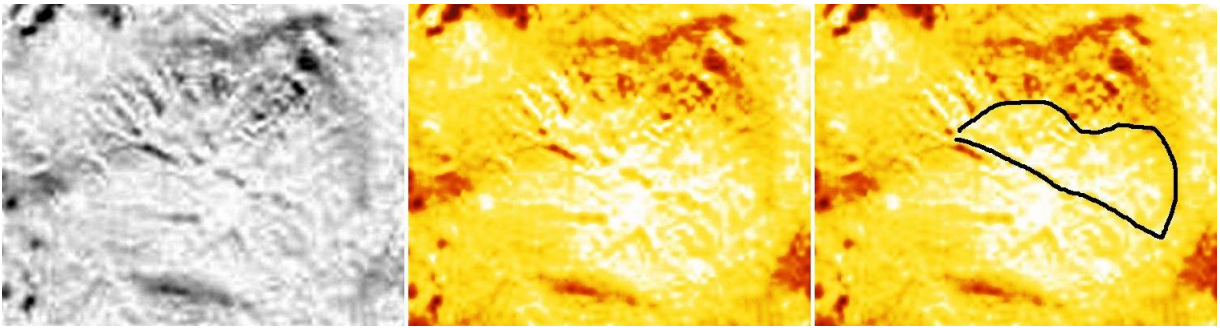

*Fig.40 The cropped central part of fig.38 features the caldera of Occator. Changed look up table of initial black and white view allows clearly see the region of white substance pile-up origin. Black line averagely separates the regions of yellow and red colors*

Thus our analyses show that the cliff mountain was ejected from the caldera, turned upside down in the flight, and landed nearby. It appeared to start with anticlockwise rotation as seen from above the surface and turned approx. 90° before landing.

The interpretation leaves no doubts that the central part of Occator is caldera's pit. This directly proves that sunlight in fig.35 falls from the left (a bit diagonally down). Hence Occator is in fact a table mountain. The same is right for several objects of compared sizes around it (e.g. see fig.28).

Occator is located in fractured area as well as Kirnis positioned in between Occator and Ahuna in fig.2(right), where fractures' directions shown by white arrow are diagonal to the planetary equator (see also 3.7). Notice shape likeness of Kirnis and Occator, as well as their similarity to a bit smaller round object with central circle, located in fig 28 approx. up-left to Occator along fracture direction at a distance twice less than that between Kirnis and Occator (cf. fig 35).

There exist other proofs of Occar's ejective origin. One of them is plausible interpretation of fig.41 (cf. fig. 32). An artificial color contrasted view allows distinguish minute surface changes. The whole picture of Occator region looks like a result of explosion with ejecta blanket thrown around, but mainly diagonally to the figure frames (darkest blue or lightest yellow in the figure). Explosion directions shown by arrows mean highly asymmetrical detonation. Therefore Occator is possible to originate the same way as Kupalo.

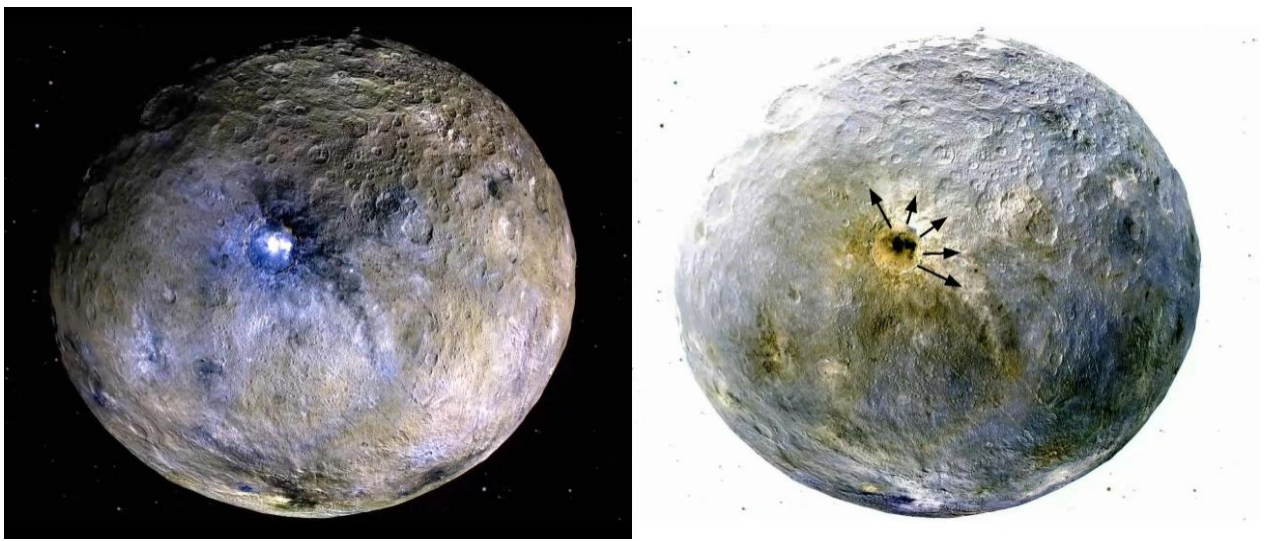

*Fig.41. Left: Occator's planetary side in false-color renderings to highlight differences of surface material. Right: the left image with inverted colors. Arrows point to feasible directions of materials' throwing out.*
*(www.photojournal.jpl.nasa.gov./catalog/pia20182)*

## 4.2 Occator under stresses

In the following we propose some speculations concerning Occator's history. The general approach is that we see the consequences of decaying volcanic activity. This means falling down the scale of eruptions and slides, ejection of sequentially smaller objects.

Pancake shaped elevation to the North of Occator appeared to slide from its surface (fig.42). Topological coincidence of features partly proves of the idea. The separation for sure happened on early stage of Occator's history. Slid cap is not the lone feature of the kind around Occator. Some of others are almost unnoticeable.

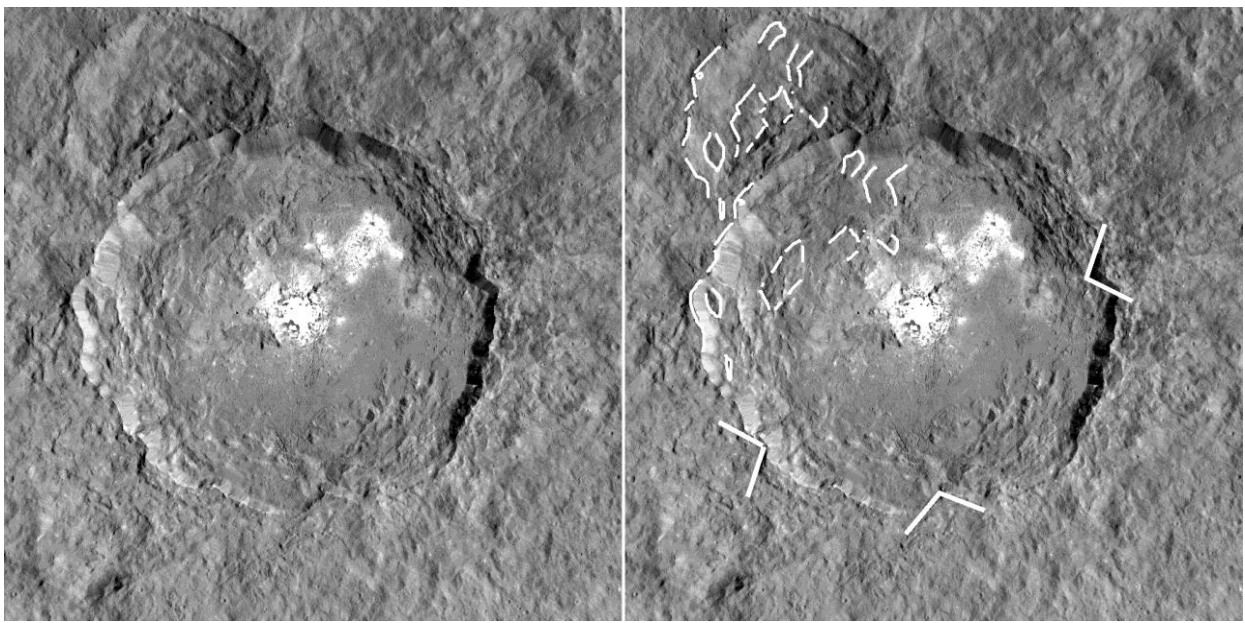

*Fig 42. Left: contrasted view cropped from pia19889. Right: the left image with coincident relief features marked. White angles mark 90° perimeter features. (www.photojournal.jpl.nasa.gov./catalog/pia19889)*

The whole Occator's picture gives the impression of strong plane tablet above weak underlying surface. The crust seems to expand, while the mountain strives to save its integrity, not without losses. That is why its perimeter is disturbed.

As said, Occator's rim is clearly seen in all its views. The southern and the eastern borders are destructed, the latter in a staircase-like manner (fig.42). The parallelism of borders and crustal surrounding features proves expansion of the underlying crust. It is clear that the eastern border is the widest so the maximal elongation acts in latitude's direction. Crustal parallel features are divided by wider spaces compared to those of the mountain (see e.g. fig. 44). Other noticeable cerean objects e.g. Dantu, Gaue, Haulani, Sekhet, Ninsar, Hamori, etc. bear the same border features. By the way, the statistics of such borders is of special interest for finding expanded regions, determination of objects' age, and history of the planet.

The crustal frictional forces along latitude elongate the mountain. According to solid state mechanics they lead to maximal shears +/- 45° to its direction. That is the reason of 90° angles of several Occator's borders (fig 42). The lowest angle points to the dyke intrusion and its bunch of fractures.

The tension differs for different latitudes. That gives rise to parallel crustal fractures along them (long white arrows fig.43). The fractures connecting the main and secondary white spots are of the same direction. The situation is analogous to Earth's transform boundaries perpendicular to Mid-Atlantic ridge which appear due to torsion. Parallel fractures are often marked by chained hills (fig.43).

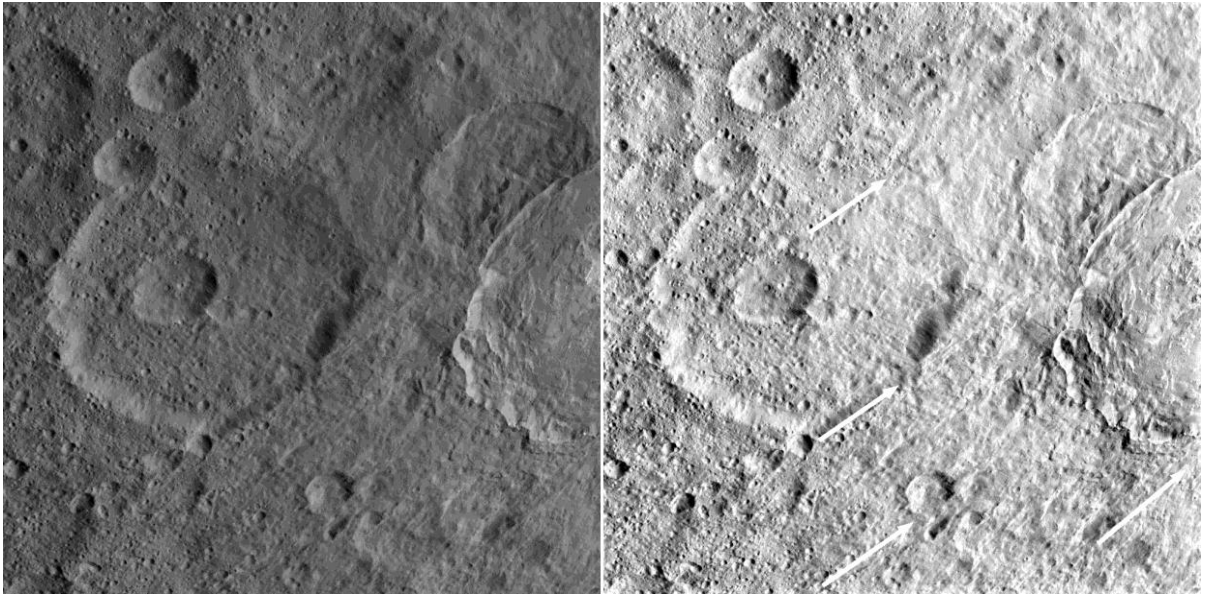

*Fig.43. Left: Half of Occator (left) and its neighbor mountain. Right: the left image with subtracted background and contrasted colors. Arrows show several transform fractures along latitudes.*
*(www.photojournal.jpl.nasa.gov./catalog/pia19991)*

Fig.44 shows the remnants of mountain and crater borders that were separated by crustal expansion. One can notice small conical hills originated as a result of ejection. We pinpointed several of them by arrows.

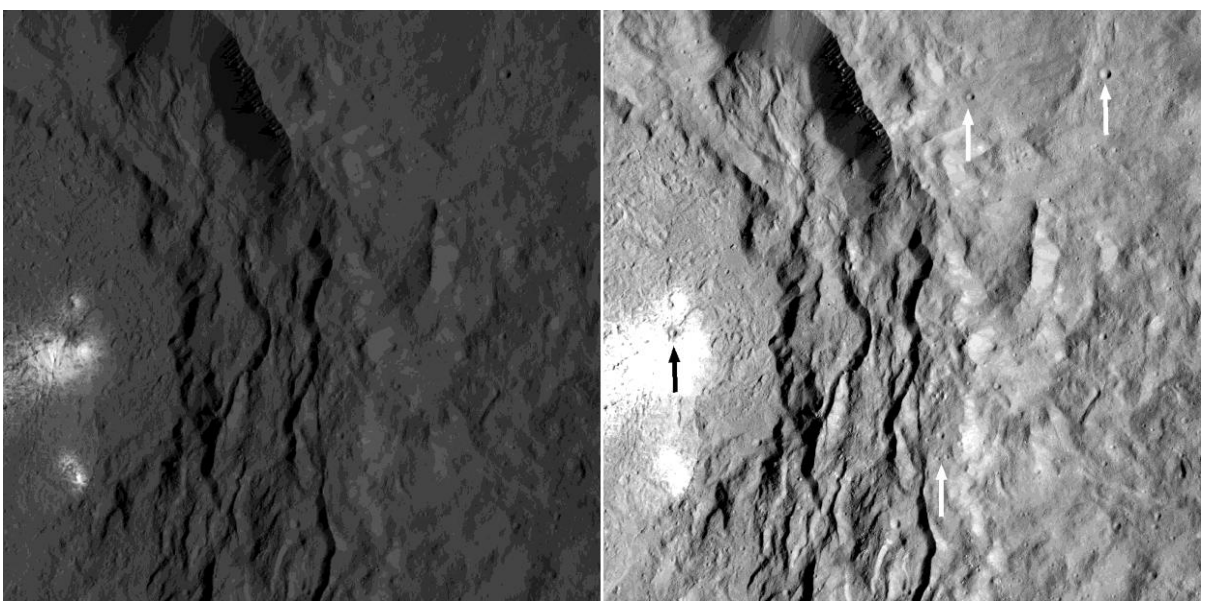

*Fig.44. Left: original view. Right: the left view contrasted. Arrows point to hills.*
*(www.photojournal.jpl.nasa.gov./catalog/pia20631)*

Now we examine details of Occator's perimeter borders (fig.45) to find out the consequences of stresses exerted by the crust on the table mountain. Magnified areas of these cliff-like slopes are shown in fig 46. It is clear that rim slope relief features are of the same nature, whether they are shadowed or not. One can see in fig.45 (middle) that there is almost no contrast between the borders, the floor, and the crust. Thus dark features mainly result from shadow contrast, not that of substance difference.

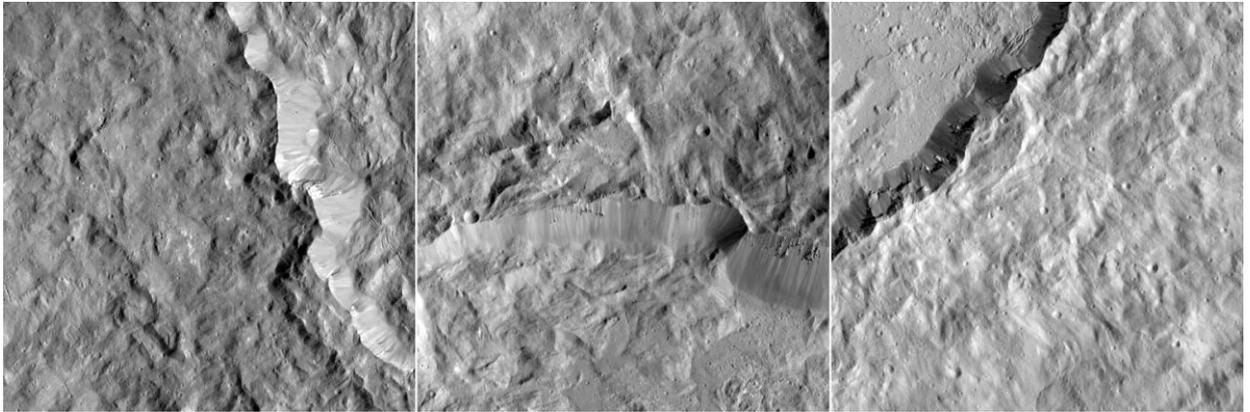

*Fig.45. Three Dawn' views of Occator's borders. Contrasts were enhanced by us. (www.photojournal.jpl.nasa.gov./catalog/pia20312, www.photojournal.jpl.nasa.gov./catalog/pia20683, www.photojournal.jpl.nasa.gov./catalog/pia20406)*

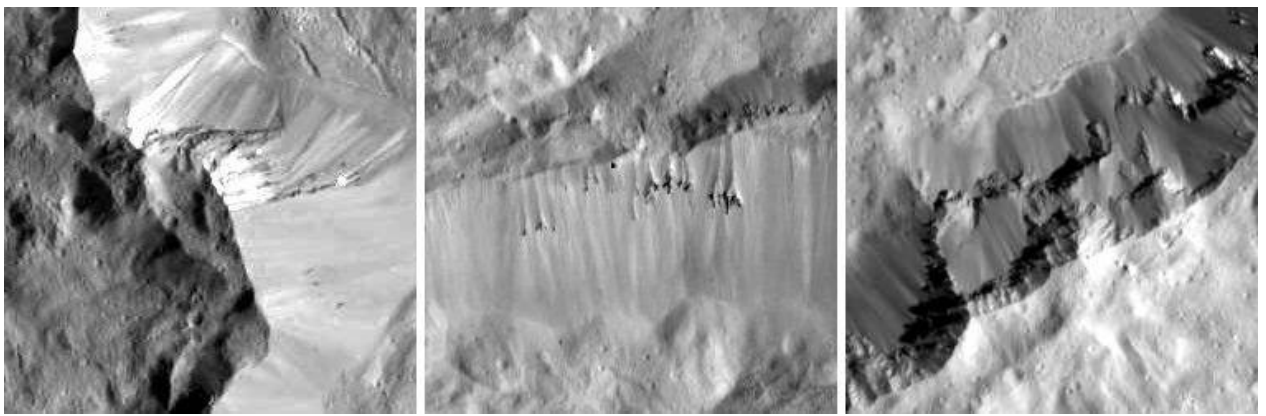

*Fig.46. Magnified crops from the every view of fig.45.*

Slope destruction features are chain pits approx. parallel to slope borders (fig.46(middle)), more elaborated horizontal remnants of cliff falls (fig.46(right)), or column-like cliff features (fig.46(left, right)). The falls are more convenient for stressed southern and eastern partly stair-like slopes. Sometimes there are separated inclusions of white substance as two bright features shown in right up angle of fig.46(right). The features discussed are genetically close to and originated from streaks.

Location of all these features appears to be determined by horizontal discontinuities between crustal layers inside the table mountain. Stresses are usually strong around them. Inner stresses inside the mountain mean its active matter stored energy and is prone to instabilities.

## 4.3 White spots and plasticity

It was shown in article [7], that there were multiple flow events inside Occator crater. They were proved by recognizable morphology. According to that work the flows were stopped due to border elevations. We showed that Occator is a mountain, but the plastic flow conclusion still holds. The process is accompanied by white substance synthesis and rock's metamorphism.

Slice by slice sliding of several Occator's layers could be explained by expansion of underlying crust which exerts tensional forces on the table mountain and inspire stresses inside it. These would lead to its differential elongation with the height. Stresses between the slices would concentrate along their discontinuity borders and are able to initiate compositional changes near them, plasticity, and even sill-like inclusions.

Different relief features of Occator's North-West quadrant outside its faculae show some similarities (figs.42, 47, segmented lines). But it is rather difficult to decipher them except for the fact that their lined borders are determined by the directions of fractures or approx. 45º and 90º to them. The impression is that upper mountain layers moved in latitude direction and then rotated.

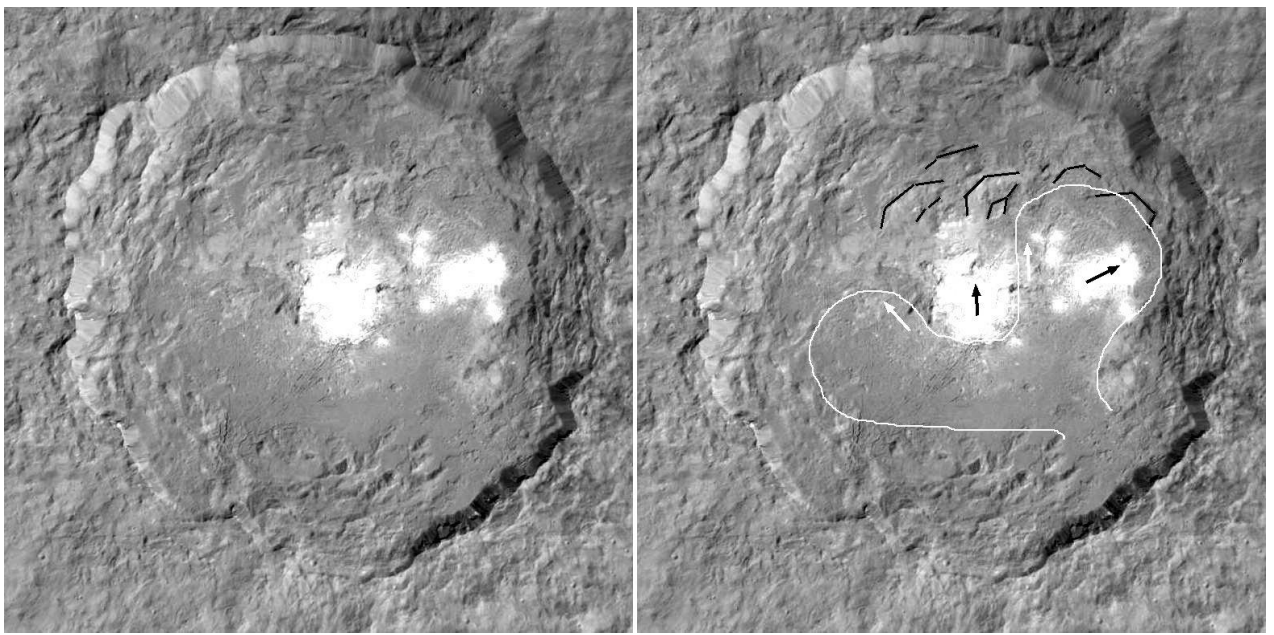

*Fig.47. Left: Occator crater image with white spots inside. It is contrasted crop from original PIA19996 view (see fig.35). Black lines depict similar relief details. White line ca. restricts u-shaped area of smooth surface relief. White arrows approx. show border directions of the area. Black arrows depict the directions of main and secondary dykes as well as their fractures.*
*(www.photojournal.jpl.nasa.gov./catalog/pia19996)*

The northern parts of black lines in fig.47 is thought of as flow margins in article [8] (fig.3 therein). We restrict the feasible flow U-shaped region with smooth relief by white line in the figure. It is interesting to note that the sizes of both sides of "U" are approx. the same. They are also equal to the size of central white spot between them. The angle between inner border directions shown by white arrows is approx. equal to the angle between the main and secondary dykes.

Besides, the size of elongated dyke remnant inside the secondary white spot is somewhat bigger than that of the main one (figs.36,37,48). The implication is that Vinalia faculae were formed in the upper mountain layer above the dyke intrusion.

Comparison of head directions and widening of the main and secondary dyke remnants allows estimate the angle of the layer rotation. It is approx. 45°. This indirectly proves the conclusion about upper layer landslide rotation.

Indirect evidences of rotation of the secondary spot layer are also round shaped relief features clearly seen in the central area of it (e.g. fig.36,48). The peripheral white spots of Vinalia faculae are approx. concentrically located relative to imaginary rotation center. These white spots are connected to local fractures and objects ejected in places of their locations (cf. fig.44 black arrow).

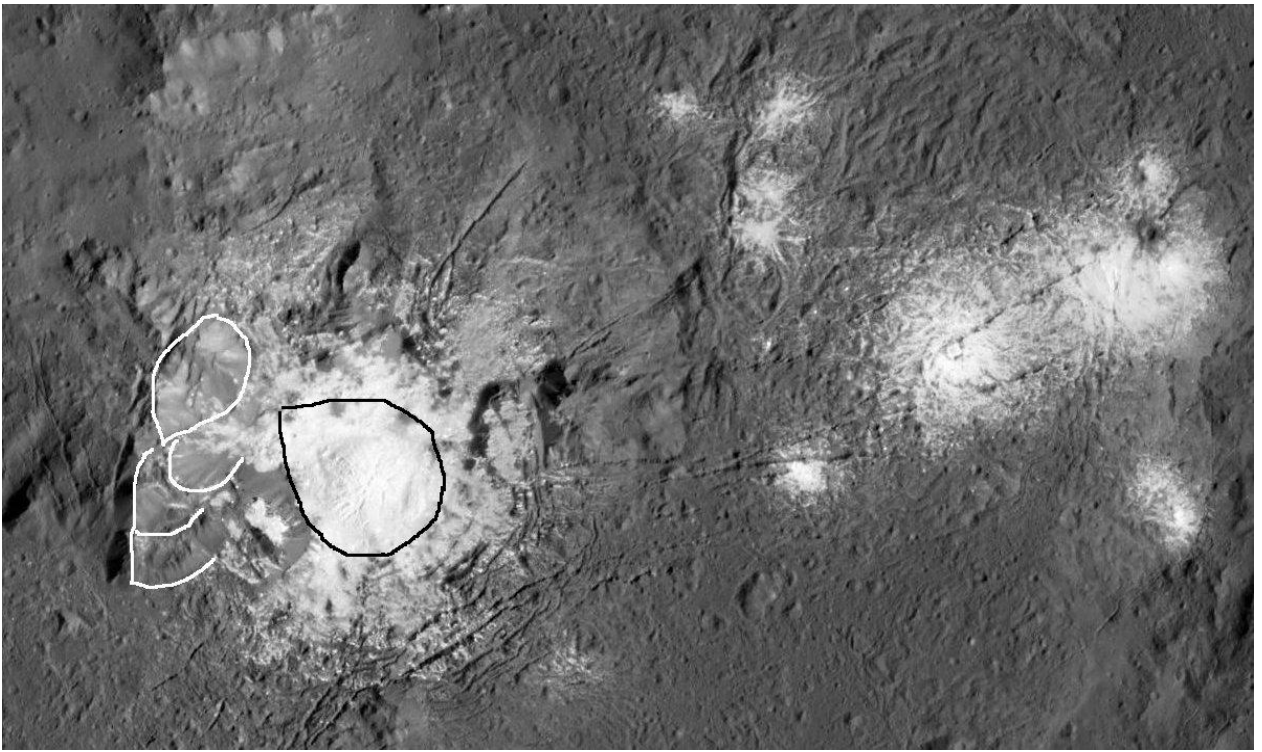

*Fig.48. A crop of white spot areas from original pia20350 view (fig.36(left)). Similar looking features are marked by white and black curves. .*
*(www.photojournal.jpl.nasa.gov./catalog/pia20350)*

Our analyses are by no means exhaustive. There are still lots of features with unclear origin on Occator's surface. For instance, we show weird leaf-like elevations by white lines in fig.48. They are possible to be ejected and/or slid. The elevations' shapes seem to coincide to the relief morphology marked by the black curve. Some additional Dawn's data, especially the spectral ones, would be of much help to completely solve the conundrum of Occator's history.

# 5 Discussion

## 5.1 Plastic rebound

Let us return to Ahuna region and discuss the problem of size differences of the crater and the mountain, as well as of the mountain table and the hill perimeter. The plasticity of the planetary crust appears to be responsible for the effect. The reasons are as follows.

Before ejection the region of future crater was under complex tensile and shear stresses. These with overall crustal plasticity led to elongation of the region. After crater formation overall rigidity of the crustal regional volume decayed. So the crustal tension lengthened the crater region in major axial direction, and shortened it in perpendicular minor one due to Poisson's effect.

Initially plastically deformed mountain contracted itself after the ejection, but has not fully returned to its previous shape due to plastic offset. It is higher for larger objects, so a rather small body has a negligible residual deformation. That is why the mountain is wider in stress direction, the smaller hill looks almost round or square shaped, and the smallest hillock seems an ideal cone in resolution of fig.2. Only in fig.15 we are able to notice its elliptical shape. By the way, this size distribution could allow the assessment of rock mechanical properties.

Internal plastic stresses due to differential with the height shortages evolve inside the mountain because of contraction after the start. The stresses alleviate cleavage along the initial crustal layers' boundaries and the formation of mountain table. Before the cleavage separation the lower and bulkier mountain parts exerted strong forces on the upper summit parts due to nonlinear with the height plastic shrinkage. The same consideration holds for inner rotations differential with the distance to rotational axes.

The plastic stress has also released around inner mountain discontinuities, for instance, strong inclusions or future table scarps. At the landing moments the abrupt linear and rotational slow down i.e. friction forces were to alleviate cleavage and the formation of curved inclusions e.g. spirals and scarps.

Our approach implies that the heights of cerean mountains are determined by the depths of crustal layers. So distribution of initial mountain heights and crater depths should correlate and have maxima at the thicknesses of regional crustal layers. This way planetary relief statistics may prove the model under consideration.

Consequently the plastic energy gained by the ejected mountain indirectly predetermines the amount and shape of large blocks and breccia material thrown out during crater formation. Deformation evolves along inherited crustal discontinuities. Different degrees of plastic deformation imply different mineral compositions and reflectivity of different mountain and crater surfaces. On the other hand, coincident parts of mountains and craters are to contain same substances, minerals, and rocks. These data are easy to acquire through spectral analysis.

## 5.2 Volcanic universality

The phenomenon of ejective orogenesis has the features of volcanic caldera formation regime. Crater streaks with their headings look like a family of Earth's ring dikes. The regime under study is possible to think of as the one that saves volcano's summit integrity or results in its breaking into several parts. The specific regime seems to be missed by geologists. Common opinion is that majority of calderas appeared as a result of rock's drop down. The reason of such conclusion is rare and weak volcanic eruptions in comparison to distant past.

An obvious example of mountain ejection and caldera formation is shown in fig.49. The round shaped object looks like volcano edifice with thrown out cone summit lying on its slope. It seems the summit sled a little down from the place of initial landing at the slope and rotated counterclockwise approximately 30° in the course of this movement.

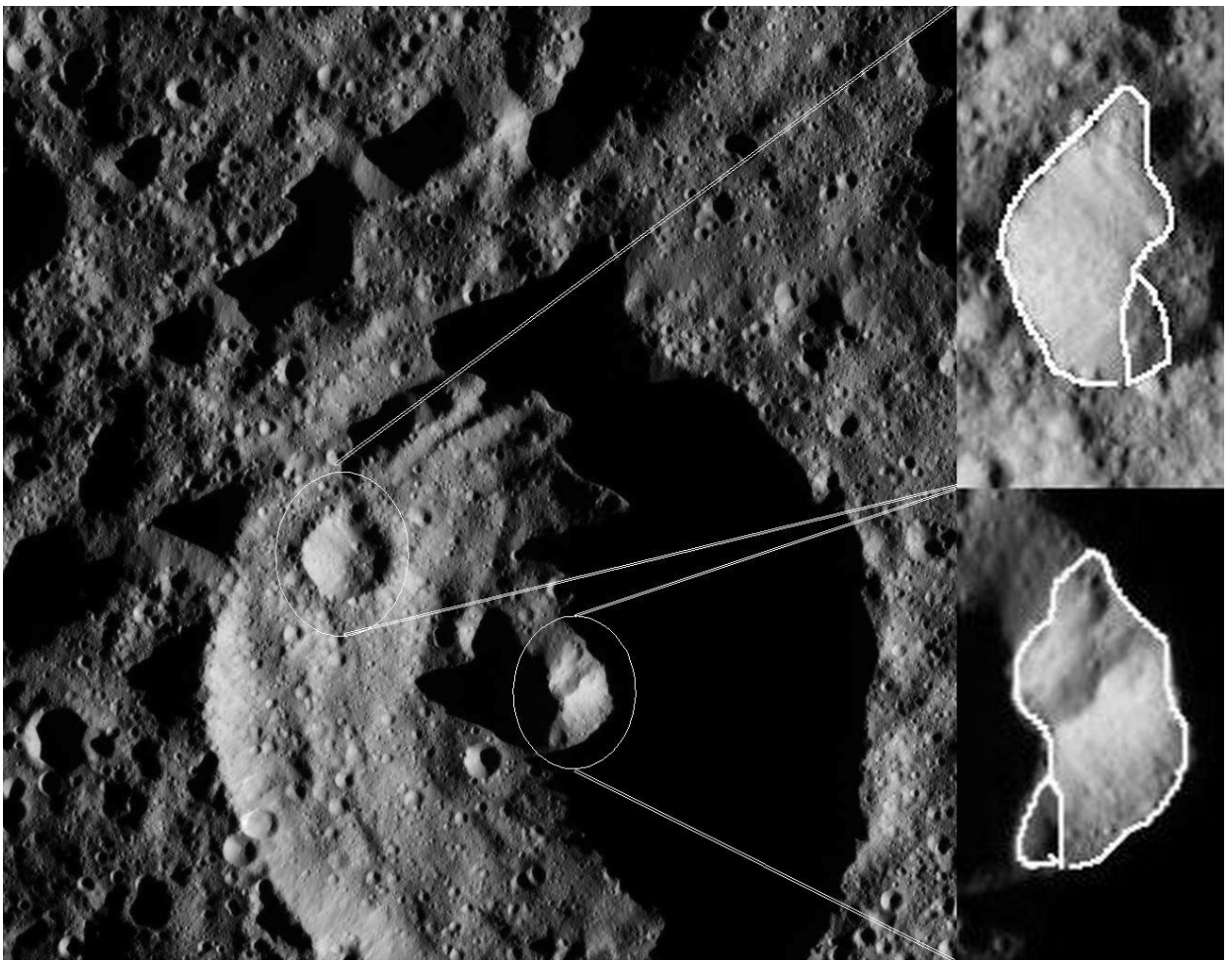

*Fig.49. NASA's pia20308 view with similar features enlarged and marked. (www.photojournal.jpl.nasa.gov./catalog/pia20308)*

Well known but purely understandable in Earth's geology phenomenon is caldera resurgence, which is an uplift of already existent caldera bottom, sometimes with small eruptions. In this case magma does not canalize into its previous vents and accumulates inside thickened caldera bottom instead. The phenomenon of resurgence is an analog of chain elevations (see 3.6).

## 5.3 Plastic zone and cone zone formation

Our approach to mountain/crater formation is universal for flat crustal surfaces, mountains, and volcanic edifices. So the first stage of the process is to be the formation of cracks and fractures under the surface. Further development will depend upon their number, lengths, depths, symmetry, which in turn are connected to the interplay of local tensions and shears as well as the nature of ambient and local materials. Therefore we have to have a number of crater/caldera/mountain/ejecta formation regimes. Resurgence seems to be the low energy one. High local energy supply will lead to complete destruction but with the shapes determined by plasticity.

Let us have a look at an object covered with ridges and steep slopes (fig.50). Image description states: "The curvilinear nature of the scarps resembles those on the floor of Rheasilvia, the giant impact crater on Vesta, which Dawn orbited from 2011 to 2012". In the light of above discussion severely curved scarps at, we think, the mountain table, are the consequences of plastic deformations. Hitherto in the central part of the mountain we have concave relief pit and a region with brighter than average substance.

It looks like the central part was severely modified and explosively thrown out of its location. Crustal fractures are directed along the arrow in fig.50(left) and perpendicular to it (see lower part of the figure). Two mountain parts (semicircles) save their integrity but look somewhat plastically moved along the fracture which direction is also given by the same arrow. Table relief geometry and inclinations imply that it is in this diagonal direction plausible explosion was to cumulate its energy. So we will try to explain the geometry of central depression and the scarps by the horizontal crack intrusion diagonal to the figure frames along the arrow direction.

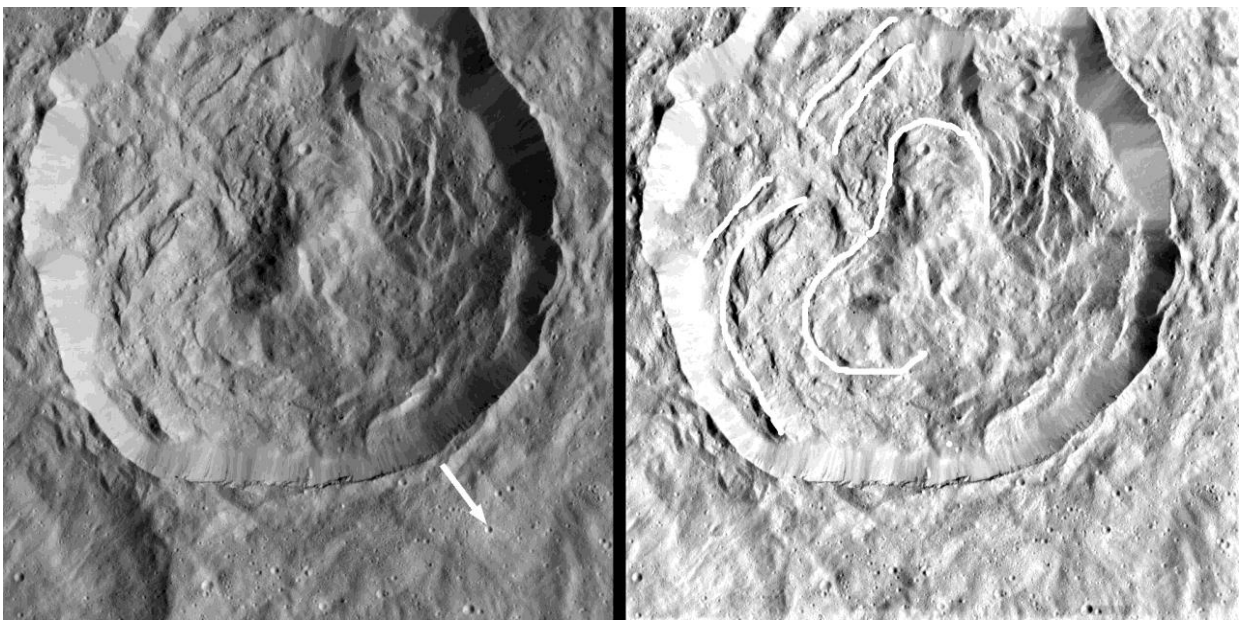

*Fig.50. 32 km wide object located at northern mid-latitudes of Ceres, just west of the larger Dantu crater. Left image is the original view, right one with its background subtracted is marked with lines. They show positions of some scarps formed due to crustal plasticity. An arrow gives the direction of plausible crack intrusion.*
*(http://www.jpl.nasa.gov/spaceimages/details.php?id=PIA20194)*

Crack evolution and existence of a plastic zone around its moving tip are well developed problems in fracture mechanics. Imagine a sheet of solid matter cut perpendicular to its surface as a model of crack extension. The zone inside the sheet, in its middle surface section has the shape of math eternity sign and the situation is described as a plain stress. The plastic zones at both surfaces of the sheet are bean shaped due to plain strain. There is continuous transition between the two mechanical regimes inside the sheet. Some examples of plastic zones are also shown in fig.51. In our above images we see the results of mixed zone mechanics.

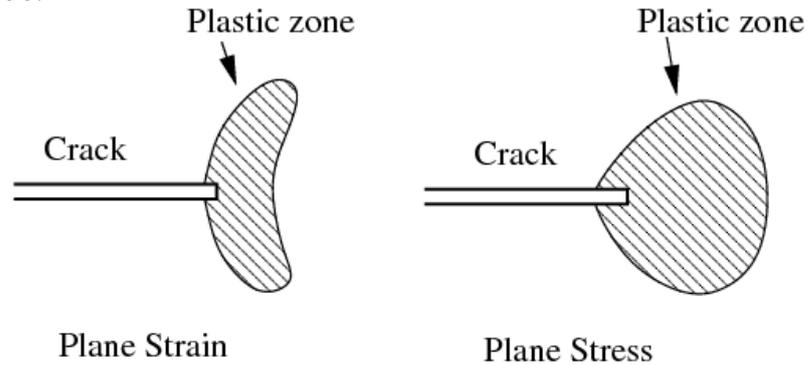

*Fig.51. The scheme of plastic zone around a crack tip in a ductile matter taken from Wikipedia's article "Fracture mechanics."*
*(http://en.wikiprdia.org/wiki/Fracture_mechanics#/media/File:PlasticZone2D)*

Another possible example of plastic zone formation is shown in fig.52. The area restricted by white bean-like loop resembles plastic deformation zone in fig.51(left). Crack intrusion directions are discussed in the figure's subscription. Arrow line intersects zone perimeter in places of approximately maximal local curvatures of the bean-like zone (cf. fig.51).

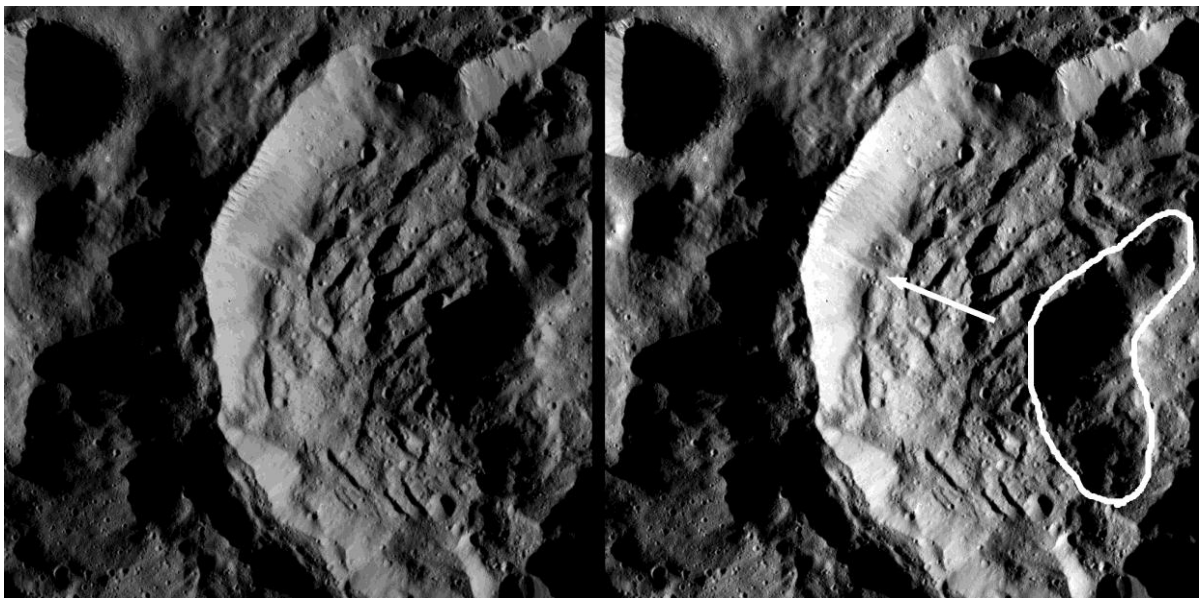

*Fig.52. Part of 40km wide Sekhet object. Left image is the original view, right one is more contrasted. The curved region marked by white line has the shape of plastic zone. Plausible fracture remnant is manifested by vertical grooves on the mountain table or by whiter than average ridge on the slope (white arrow)*
*(http://photojournal.jpl.nasa.gov/catalog/PIA20390)*

Conical shapes of explosively ejected mountains and their craters are possible to be explained by vertical crack intrusions and local fault formation. The shape of plastic zone in fig.51(right) is that of a vertical mountain sectioning by half. In the logic of fracture mechanics shear stress between cone halves or parts alleviates plastic zone formation. As well it is the reason of ejected cone rotation. Zone eternity-sign shape will lead to the origin of twin craters juxtaposed along their fracture line. Horizontal crustal tension around future crater will lead to maximal shear stress, directed 45° to surface plane. That predetermines slope 45° angles and close to 90° summit angles.

Thus mountain rim shapes in figs. 51, 52 are possible to be explained by vertical crack intrusions leading to explosive orogenesis. For Ahuna's case the triple branching remnants of intruded and healed cracks are clearly seen (e.g. see fig. 6). In Earth's conditions the remnants of the kind are quite usual. Remember, for instance, the straight groove of Mount Kailash in Tibet region.

Discussed plastic shapes are familiar to geologists, but for objects of much smaller scales. Those are the shapes of tektites, metamorphic rigid substances forming far from equilibrium. The formation conditions are huge transient temperatures and shock pressures (10-100GPa) of Earth's impact craters and underground nuclear tests. In case of higher morphing energies and pressures more round shaped and spherical tektite bodies of centimeter sizes are formed. To date there is no unanimous theory of those glassy bodies' formation. Theoretical proposals include very exotic approaches such as that of nuclear physicist Matest Agrest.

The phenomenon of ejective orogenesis bears well known features. Conical rocks of meter scales and less are known on Earth. Conical horse tail shapes in section is the feature of rock shatter cones, which are shock metamorphic structures. Cones are usually found in groups under surfaces of so called impact craters or in places of underground nuclear tests. On crystallographic scales the cone analogs are crystal twins. Along with gliding dislocations they arise in crystal lattices due to plastic deformations.

The mechanism of cliff shape formation of shatter cones is still under study. Some authors think cones evolve in conditions of severe pressures and plasticity. Our analyses prove this approach and expand it to scales orders of magnitude larger. Close geometrical resemblance of twins, cones, cliffs and ejected mountains makes us think of unified shock mechanism of their origin. It implies the initiation of transient tension, development of shear stress, formation and lengthening of disjunctions and cracks, and gliding of material's layers. There appears to exist a continuous row of pressure and energy evolution states from shatter cone bunches, through conical in section, bean and eternity-sign-shape tektites, to separated spherical and oval ones, which spread over vast Earth's regions (e.g. North-African deserts).

According to above considerations an ejected mountain as a whole is usually stronger than crust around its crater. It is plastically modified along the surfaces predetermined by mountain inner structure. If the three dimensional strain is of cone in cone type, the cones of sequentially smaller sizes repeat the overall mountain symmetry. The ca. vertical mountain section would manifest the horse tail or wedge in wedge structural geometry. That's why mountains should destroy and weather themselves layer by layer, the same way as cliffs do, i.e. along slip cone surfaces and horizontal planes, which are remnants of crustal discontinuities. That is just the case.

## 5.4 Explosive nature of ejection

How could it be, that mountain as large as, say, thousands of cubic kilometers is thrown tens of kilometers up? The only reason we would imagine is an extreme explosion, the one that happened almost simultaneously all over conical crater surface and ejected mountains to fly without destruction. The one that had the energies compared to the most powerful known on Earth and resulted in the synthesis of brighter material.

Size of asteroid Ceres is approximately thirteen times less than the Earth's one, the density is almost three times less, and the gravity is only 2,7% of that of the Earth. So for the case of Ahuna the time of flight estimation making use of classical free fall formula gives approx. dozen of minutes.

For conical explosion the forces exerted on a mountain will depend on the widths and inclinations of crater slopes. The steeper the slope the more is the horizontal force and the less is the vertical one, i.e. the steepness difference means unbalance of ejective forces. In case of Ahuna the upper crater slope is steeper than the opposite one (e.g. fig.2). That is why the mountain moved down (in relation to the figure frames). The conclusion is proved by the topography of Ahuna crater [9]. Upper part of the crater slope is covered with regolith-like debris with the elevation tongue thrown out half of crater diameter. In case of Kupalo the shadowed mountain slope looks several times steeper than the bright one (e.g. fig.17), so the flip torque of highly asymmetrical explosion sided the mountain up just at its birthplace location.

Saving mountains' integrities after falling down looks like counterintuitive result. The physical reasons of this are small gravity of Ceres and plasticity of its crust, which allows the thrown out objects to bear some qualities of viscous liquid droplets and partly adopt their inner structures to the energies of vast ejection explosions. Besides, the gravity center position of the starting mountain is always lower than that of its final location, so the maximal free fall velocity is less compared to the starting one. Hence the possibility of ejected mountains' destructions in plain regions is mainly determined by starting accelerations.

To estimate the upper margin of ignition time we divide the mountain's size by the velocity of sound. The time is less than 0.1sec. The explosive waves from ignition point should run around the buried cone and meet somewhere in opposed region. So the power release here is to be the highest. There should be at least one, so to say, singular region in opposition to ignition area.

For the case of fig.11 the hypothetical region is pinpointed by short black arrow. There are discontinuities of perimeter border in places like this due to destruction of the fallen mountains. The reason seems to be the local strengthening (up to destruction) and severest explosive modification of regional crystalline and chemical structure, sometimes with the formation of large amounts of brighter substances. The area and its vicinity appear to form out of the most plastically deformed rocks. Notice the biggest curvature of local mountain rim which inherits the position of local crustal fracture. We suppose that opposite ignition fuse was also located somewhere inside it.

The same arguments are applicable to other above studied objects (see e.g. fig. 23 black arrow). They all have at least one singular region. Those regions seem the first mountain parts to start and to fall. Whether they are the first to destruct depends upon mountain's strengthening and regional plasticity.

### 5.5 Transients and planetary facts

Now we discuss some transient phenomena of Ceres. According to Dawn's team a surprising observation came from gamma ray and neutron spectrometer [10]. The instrument detected bursts of energetic electrons. This unexpected data are not yet fully understood, hypotheses on the matter are testing now. To our opinion the bursts are possible to be explained by explosive events in the course of contemporary small scale ejective orogeneses.

Slow chemical processes ignited by crustal plastic stresses may explain light hazes above craters of Ceres. Those explain occasional water vapor exhaust seen in Hershel telescope [4] and Ceres' atmospheric composition changes [1] . Weak and slow processes are possible to be modulated by crustal stress periodical changes which could be the consequence of planetary rotation-vibration coupling. Sun's illumination and heating also may lead to destruction of ejected substrates. So the diurnal periodicity in different data on the dwarf planet is quite feasible. Tiny transient exospheres with water vapor as main component are also plausible result of crustal modifications and exhausting.

Current analyses allow explain some other planetary facts. Depths of floor fractured craters on Ceres are several hundred meters (up to one kilometer) shallower than others [8]. If they are in fact mountains then this is explained by faster destruction of their surfaces, because larger objects are more prone to it due to accumulation of layer misfit. The same may be the reason of lack of big craters (>280km) on Ceres. Areas of low crater density are associated with major craters such as Urwara, Yalode, Kerwan [8]. The explanation is that their surfaces are mechanically relaxed.

In general borders of craters are usually diffuse and less clear than those of mountains (e.g. see figs.2,5,7,17). Probably craters are more prone to degradation because they are not deeply modified, while thrown out bodies become stronger due to plasticity occurring in them. Another plausible reason is the thick regolith cover consisting of explosive debris. These substances should contain different rocks due to explosive modifications. The border difference would simplify crater/mountain visual search.

Bright substances are dispersed all over the surface of Ceres and often look like round blocks of sizes small compared to craters. On crater borders brighter colored lines are also formed as streaks. Less pronounced or sometimes degenerated into lined blobs are they on mountain slopes. We conclude that bright substances of different weathering stage are either the remnants of large ejected objects' destruction or formed in their places of location due to smaller scale ejective orogeneses. The scale of planetary geologic processes has been diminishing.

# 6 Conclusions

We studied the phenomenology of mountain and crater formations on dwarf planet Ceres. The detailed analysis of relief features allows us to prove that a tall mountain in Ahuna region was ejected rotated out of Ceres' surface with the simultaneous formation of its crater due to conical explosive cleavage of the planetary crust. The mountain flipped, landed nearby after dozen minutes of flight, and its upper part separated to form the smaller table mountain.

The newly described phenomenon of ejective orogenesis and subsequent landslides were shown to explain formations of other relief objects of Ceres. Ejected mountains are proved to be the plastically modified parts of the planetary crust. They save some of its properties, which later may determine their future, e.g. destruction. There is a special regime of the orogenesis, when ejected mountain goes down above its crater thus masking it.

It was shown that famous for its white spots object, known as Occator crater, actually is a table mountain. Inside it we found ejected smaller cliff-like mountain with the shapes coincident to its caldera and partly covered with white substances. The whole edifice of Occator demonstrates decaying volcanic activity with plastic flows and sliding. The white spots were also synthesized in the course of caldera formation during ejective orogenesis. Therefore the same phenomenon is shown to be responsible for plane and volcanic crater formation.

The reasons of the orogenesis are tensions and shears natural to the planetary crust of expanding Ceres. Under those stresses the crust with inherent stress concentrating cracks and fractures locally becomes an active media prone to instabilities and phase transitions. The locations of ejected objects are connected to the locations of these disjunctions, which in turn are hierarchically determined by the global net of planetary fractures and discontinuities. Geometrically regular planetary crustal thickening, sometimes periodical, is possible to be explained by the same phenomenon, when the progress of ejective orogenesis ceases at its initial stages.

It is common belief in geology that mountains grow due to compression, the growth is very slow and takes a lot of time large compared to a human lifespan. So it is challenging to imagine quick to say nothing of explosive appearance of mountains. In this article we return to cataclysmic ideas of geology founder James Hutton which were erased by his successor Charles Lyell.

From the above analyses the existence of the phenomenon of ejective orogenesis is obvious, but the features and parameters are to be investigated further. Our preliminary inquiry proves that natural for Ceres phenomenon is the ubiquity among all rigid celestial bodies.